\documentclass[twocolumn, times, tighten, twocolappendix, astrosymb]{aastex63}
\usepackage{hyperref}
\usepackage{amsmath}
\usepackage{amssymb}
\usepackage{graphicx}
\usepackage{color}
\usepackage{longtable}
\usepackage{lineno}
\usepackage[utf8]{inputenc}
\DeclareUnicodeCharacter{2212}{-}
\usepackage{calc}

\newcommand{\kms}{kms$^{-1}$}
\newcommand{\teff}{$\log(T_{\rm eff})$}
\newcommand{\vsini}{$v\sin{i}$}

\newcommand{\av}{$A_{\rm V}$}
\newcommand{\logg}{$\log(g)$}
\newcommand{\hst}{\textit{HST}}

\newcommand{\zsun}{$Z_{\odot}$}
\newcommand{\msun}{$M_{\odot}$}
\newcommand{\rsun}{$R_{\odot}$}
\newcommand{\beast}{\texttt{BEAST}}

\newcommand{\disappear}[1]{}

\begin{document}

\title{Unveiling Massive Main-Sequence Stars in Sextans A through Panchromatic Photometry}

\author[0000-0003-3747-1394]{Maude Gull}
\affiliation{The Observatories of the Carnegie Institution for Science, 813 Santa Barbara Street, Pasadena, CA 91101, USA}
\affiliation{Department of Astronomy, California Institute of Technology, Pasadena, CA 91125, USA}
\affiliation{Department of Astronomy, University of California, Berkeley, Berkeley, CA 94720, USA}

\author[0000-0002-6442-6030]{Daniel R. Weisz}
\affiliation{Department of Astronomy, University of California, Berkeley, Berkeley, CA 94720, USA}

\author[0000-0003-1680-1884]{Yumi Choi}
\affiliation{NSF National Optical-Infrared Astronomy Research Laboratory, 950 North Cherry Avenue, Tucson, AZ 85719, USA}

\author[0000-0002-7502-0597]{Benjamin F. Williams}
\affiliation{Department of Astronomy, University of Washington, Box 351580, U.W., Seattle, WA 98195-1580, USA}

\author[0000-0003-0394-8377]{Karoline M. Gilbert}
\affiliation{Space Telescope Science Institute, 3700 San Martin Dr., Baltimore, MD 21218, USA}
\affiliation{The William H. Miller III Department of Physics \& Astronomy, Bloomberg Center for Physics and Astronomy, Johns Hopkins University, 3400 N. Charles Street, Baltimore, MD 21218, USA}

\author[0000-0002-1264-2006]{Julianne J.~Dalcanton}
\affiliation{Center for Computational Astrophysics,
Flatiron Institute,
162 Fifth Ave,
New York, NY, 10010, USA}
\affiliation{Department of Astronomy,
University of Washington,
Box 351580,
Seattle, WA, 98195, USA}

\author[0000-0002-6871-1752]{Kareem El-Badry}
\affiliation{Department of Astronomy, California Institute of Technology, Pasadena, CA 91125, USA}

\author[0000-0001-8867-4234]{Puragra Guhathakurta}
\affiliation{Department of Astronomy \& Astrophysics, University of California, Santa Cruz, 1156 High Street, Santa Cruz, CA 95064, USA}

\author[0000-0002-8937-3844]{Steven R. Goldman}
\affiliation{Space Telescope Science Institute, 3700 San Martin Dr., Baltimore, MD 21218, USA}

\author[0000-0001-5538-2614]{Kristen B.~W. McQuinn}
\affiliation{Space Telescope Science Institute, 3700 San Martin Drive, Baltimore, MD 21218, USA}
\affiliation{Department of Physics and Astronomy, Rutgers, The State University of New Jersey, 136 Frelinghuysen Rd, Piscataway, NJ 08854, USA}

\author[0000-0002-1445-4877]{Alessandro Savino}
\affiliation{Department of Astronomy, University of California, Berkeley, Berkeley, CA 94720, USA}

\author[0000-0003-0605-8732]{Evan D. Skillman}
\affiliation{Minnesota Institute for Astrophysics, University of Minnesota, 116 Church Street South East, Minneapolis, MN 55455, USA}

\begin{abstract}
We present a study of the metal-poor ($\sim6\%$\zsun) massive ($>8$\msun) main-sequence star population in the star-forming dwarf galaxy Sextans A. By modeling near-UV to near-IR photometry of individual stars using the Bayesian Extinction and Stellar Tool (\beast) we infer stellar parameters such as effective temperature, luminosity, and initial mass. We identify 867 massive main-sequence star candidates (present-day mass $>8$\msun\ and surface gravity $>3.7 dex [cgs]$) with a plausible spectral energy distribution (SED) fit, $\sim 500$ of which show a probable SED fit. Comparisons to spectral types of existing observed spectra are consistent with the \beast-derived stellar parameters, with most discrepancies explained. We identify 292 OBe star candidates through IR photometric signatures and find lower-limit OBe fractions of $15\%$ for M $\gtrsim$ 8 \msun, $23\%$ for M $\gtrsim$ 15 \msun, and $17\%$ for M $\gtrsim$ 20 \msun. We find 57 OB associations and that $24–28\%$ of massive stars are isolated (distance to nearest massive star $>28$ pc). We discuss six likely runaway candidates (suggested velocities of $\sim 50-340$ \kms) not clearly associated with any major star-forming complexes. Lastly, we predict Lyman continuum (LyC) escape fractions of $f_{\rm esc}=0.27–0.76$ across the star‑forming regions and a global value of $0.35–0.71$ by assuming low overall extinction and a range of porous geometries, indicating efficient leakage of ionizing photons. Future spectroscopic follow‑up and resolved ISM studies will refine these constraints and solidify Sextans A as a benchmark for studying massive‑star evolution and feedback at extremely low metallicity.
\end{abstract}

\keywords{stars: massive – stars: metal-poor -- galaxies: individual: Sextans A – galaxies:
stellar content}

\submitjournal{ApJ} 

\section{Introduction}
Across the universe, metal-poor massive stars influence astrophysical phenomena and dominate the light of metal-poor star-forming galaxies beyond the Local Group. The stellar physics, evolution, and final fates of massive stars can change drastically as a function of mass and metallicity, significantly impacting their environment and host galaxy \citep[e.g.,][]{Massey03,ekstrom_massive_2021,Eldridge21,marchant_evolution_2023}. However, due to the dearth of spectroscopic data in the lowest-metallicity regime, most of our theoretical stellar models remain loosely empirically constrained. Significant uncertainties thus persist in our understanding of massive stars, especially at low metallicities.
As a result, the interpretation of astrophysical phenomena tied to their stellar evolution and final fate remains challenging. Examples of such results are the recent findings in the high-redshift Universe by JWST  \citep[e.g.][]{bunkerJADESNIRSpecSpectroscopy2023,cameronNitrogenEnhancements4402023,senchynaGNz11ContextPossible2024,castellanoJWSTNIRSpecSpectroscopy2024,toppingMetalpoorStarFormation2024,toppingDeepRestUVJWST2024}, the gravitational wave observations by LIGO  \citep[e.g.][]{Abbott16,Klencki18,Mandel22,marchant_evolution_2023}, or various transient observations \citep[e.g., supernovae, gamma ray bursts][]{Woosley06,smartt_progenitors_2009,chevalier_common_2012,metzger_luminous_2022}.

Significant efforts have been spent on studying low-metallicity massive stars in the Local Group (LG) through both photometric and spectroscopic observations. For example, the Small Magellanic Cloud (SMC) has been studied extensively due to its proximity \citep[$D\sim 62$~kpc;][]{Graczyk20} and low metallicity ($Z \sim 20 \% $~\zsun) (e.g., VLT-FLAMES Survey; \cite{Evans06, mokiem06, trundle07, Hunter08, hunter09, Ritchie10, dunstall11}, RIOTS4; \cite{lamb16, oey18}, ULYSSES; \citet{romanduval20, vink23, Sana24,sander24,martins24, Ramachandran24, BerniniPeron24}, BLOeM; \citet{Shenar24, Bestenlehner25, Patrick25, Bodensteiner25, Villasenor25, Britavskiy25}. However, SMC metallicity is not sufficiently metal-poor to serve as a proxy for low metallicity environments ($Z < 0.1 $\zsun) \citep{eldridge_new_2022}. Physical processes with strong metallicity dependency \citep[e.g., star-formation, stellar evolution, binary evolution, stellar feedback,][]{Conroy09,Mannucci10,Smith2014,marchant_evolution_2023} remain empirically unconstrained. 

In the past decade, tremendous spectroscopic efforts have been made to observe large samples of massive stars at sub-SMC metallicities.\citep[e.g.,][]{camacho16,evans19,garcia19a,telford21, Ramachandran21, Lorenzo22, Gull22, telford24,Mintz25}. However, due to distances to most of these star-forming sub-SMC metallicity galaxies, these massive stars are faint and challenging to observe in large quantities. Additionally, metal-poor massive stars are rare locally compared to their high-metallicity counterparts. Our current spectroscopic capabilities do not allow us to study more than a couple of dozen stars at a time, making it challenging to study massive stars as an entire population. Alternatively, using UV-optical-IR photometry and modeling the broadband spectral energy distribution (SED) of resolved stars provides an observationally more efficient way of studying a significant population of stars as an ensemble \citep[e.g.,][]{Calzetti15,Dalcanton12,Williams21,Murray24,Gilbert25}.   

The massive star population of Sextans A has been studied photometrically \citep[e.g.,][]{Massey07b,Massey07,Bianchi12,Britavskiy14,Britavskiy15,Flores26}. 
Early studies of massive stars used ground-based photometry that limited extensive analysis of the population due to a combination of the limited (1) spatial resolution ($\sim 0.27''$), (2) wavelength coverage (B band $\sim 3600 \mathring{A}$), and (3) shallow photometric observations (V $> 22.5$ mag and U  $> 19.6$ mag). These factors particularly inhibited the study of crowded regions, limiting the analysis to the brightest stars in the galaxy. Significant progress was made through the HST-based photometric study of \citet{Bianchi12}, which focused on the young hot stellar population of Sextans A. The study fit spectral energy distribution (SED) models to $\hst$ WFPC2 observations of 571 stars and demonstrated that UV imaging, $\hst$ spatial resolution, and SED fitting, rather than color-magnitude diagram (CMD) fitting, could broadly characterize hot young stars. Despite using the state-of-the-art tools available at the time, SED fitting was limited to iterating over a small grid of discrete values. It did not include artificial star tests, which capture various biases in fitting panchromatic photometry (e.g., flux calibration and crowding).  

Here, we present an in-depth panchromatic photometric study of metal-poor massive stars in Sextans~A ($Z\sim6$\% \zsun, $D\sim1.3$~Mpc; \citealt{skillman89,Dalcanton09}). Sextans A is known to host massive stars \citep{DohmPalmer02,Bianchi12,camacho16}, but none of the previous photometric studies published the associated stellar parameters (e.g., effective temperature, mass, surface gravity, etc.). Most recently, \citet{Flores26} published a catalog of massive stars with basic stellar parameters (temperature , age, surface gravity, and mass) determined by applying a finite element interpolation of the stellar tracks to the ultraviolet CMD. Spectroscopically, \citet{Lorenzo22} published an extensive qualitative analysis of optical spectra of OB stars in Sextans A, two stars were observed as part of ULYSSES (UV observations), and an in-depth spectroscopic medium-resolution optical study of individual stars was only recently published \citep{telford24}. However, only one star currently has stellar parameters provided \citep[labelled S3;][]{telford24}. This panchromatic photometric study will provide detailed stellar parameters derived from photometry for Sextans A. The aim is to characterize massive stars and their role within their host environment while further establishing that UV-opt-IR photometry can study these properties beyond our spectroscopic reach.

The paper is organized as follows. In \S \ref{sec:data}, we describe the multi-wavelength \hst\ observations and data reduction. In \S \ref{sec:beast}, we detail the process of measuring stellar properties from the \hst-based spectral energy distributions (SEDs). 
In \S \ref{sec:mstar} and \ref{comp}, we present our SED fitting results, provide a detailed discussion of them, and compare them to the literature. In \S \ref{sec:bestars}, we discuss the OBe star candidates: how they are identified and their fraction. In \S \ref{sec:spati}, we identify OB associations and field stars. In \S \ref{sec:gaslit}, we compare our stellar populations study to studies of atomic and ionized gas, as well as far-UV (FUV) photometry. Lastly, in \S \ref{escapef}, we predict the escape fraction of Sextans A. In the appendix, we highlight the importance of UV filters (Appendix \ref{sec:UV}) and the uncertainties due to fixing parameters in the fitting (Appendix \ref{sec:fix}), and discuss the limitations of our current SED-fitting for massive binaries (Appendix \ref{sec:singletracks}).

\begin{figure*}[th!]
\includegraphics[width=\textwidth]{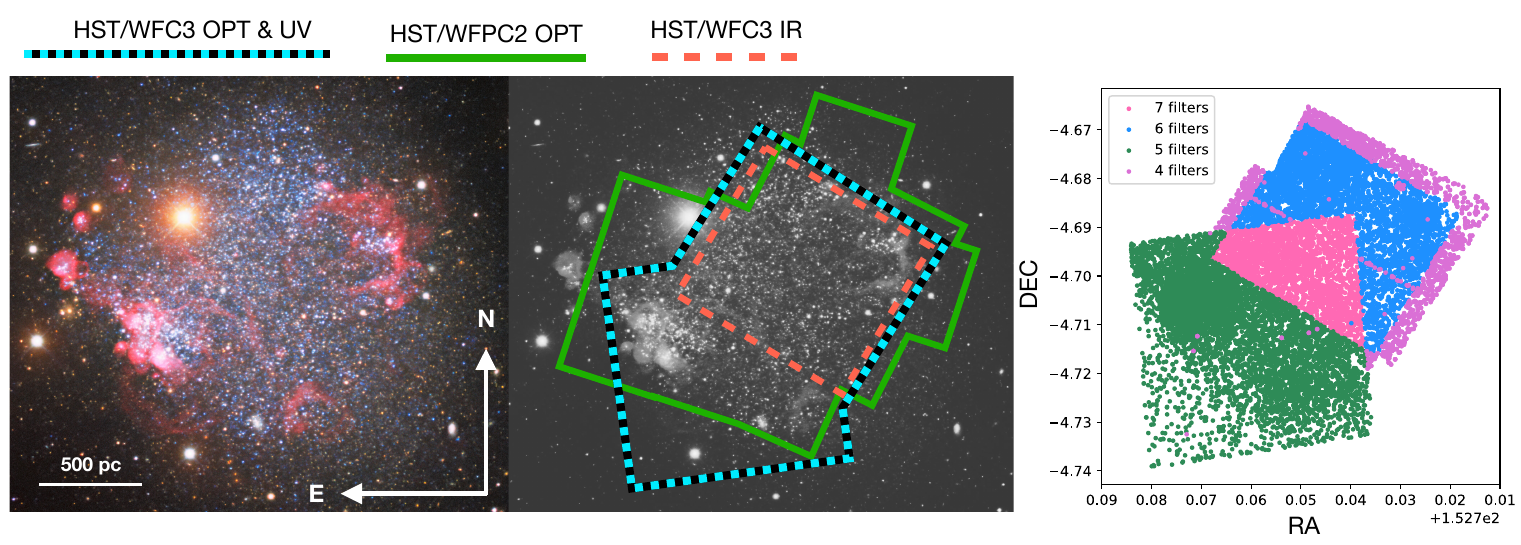} 
 \caption{Optical image (left: color, right:grayscale) of Sextans A obtained with the Nicholas U.\ Mayall 4-m Telescope at Kitt Peak National Observatory. Left: We indicate the directions of North and East, along with a distance scale of 500 pc. Middle: Footprints of the HST imaging produced by the LUVIT Survey \citep{Gilbert25} are overlaid on the optical image. The turquoise footprint shows the HST UV observations, the green and black footprint shows the HST optical observations, and the red footprint denotes the HST IR footprint. (credit: KPNO/NOIRLab/NSF/AURA Data obtained and processed by: P. Massey (Lowell Obs.), G. Jacoby, K. Olsen, \& C. Smith (AURA/NSF) Image processing: T.A. Rector (University of Alaska Anchorage/NSF NOIRLab), M. Zamani (NSF NOIRLab) \& D. de Martin (NSF NOIRLab)
Right: Coordinates of resolved sources within the footprint, color-coded by the number of observed sources.} \label{pos}
\end{figure*}

\section{HST Photometry} \label{sec:data}

To characterize massive stars in Sextans A, we use archival UV, optical, and IR \hst\ imaging. The footprints for our datasets are shown in Figure~\ref{pos}. 

We use the {\it HST\/} data collected as part of the Local UltraViolet and Infrared Treasury (LUVIT) data products. The LUVIT data collection and reduction were carried out separately using DOLPHOT \citep{Dolphin00,DOLPHOT}, and we refer the reader to the survey paper \citet{Gilbert25} and \citet{Gull22} for further details. We briefly summarize the reduction here. The footprints for our datasets are shown in Figure~\ref{pos}.

The LUVIT data for Sextans A include \hst\ UV, optical, and NIR broadband imaging. Optical imaging in the WFPC2 F555W and F814W filters acquired as part of GO-7496 \citep[PI E. Skillman; ][]{Dohm-Palmer02}, and supplemented by new WFC3/UVIS F225W, F275W, F336W, F475W and F814W (GO-15275; PI K. Gilbert \& GO-16104; PI J. Roman-Duval) and WFC3/IR F110W and F160W (GO-16162 (PI M. Boyer) imaging. To summarize the filter coverage, we provide Table \ref{filtercombo} and a mapping to Sextans A in Figure \ref{pos}. We note that at most 7 filters are used in the fitting, although the majority of stars will have either 5 or 6 filters used in their SED fit. 

To reduce the {\it HST\/} data, the photometric reduction pipelines for \hst\ simultaneous $N$-band photometry \citep{Williams14, Williams21} were utilized and improved \citep[see][for details]{Gilbert25}.

For a given reference image footprint, DOLPHOT reports measurements of point spread function photometry for every object detected in any exposure. The measurements include the flux, flux uncertainty, magnitude in the Vega system, SNR, and various photometric quality metrics. Quality cuts are applied to the DOLPHOT output catalog of all objects identified: 1) SNR $\geq$ 4 for all; (2) crowding $<$ 1.3 (WFC3/UVIS), $<$ 2.25 (WFC3/IR, ACS/WFC), and $<$ 2.00 (WFPC2); and (3) square of sharpness (SHARP$^2$) $<$ 0.15 (WFC3/UVIS, WFC3/IR), $<$ 0.2 (ACS/WFC) and, $<$ 0.25 (WFPC2), producing the ``GST'' catalog. These quality cuts were applied independently to each band. Additionally, objects associated with or impacted by diffraction spikes are removed using the median filter to identify affected regions and culling objects belonging to those regions \citep{Gilbert25}. We limited our stellar SED modeling to the bright targets of our interest by applying the following cut: $F475W < 24$ mag. 

Artificial star tests (ASTs) were run for 234499 stars and are used to compute the photometric accuracy, precision, and completeness. See \citet{Gilbert25} for more details.

\begin{figure*}[ht!]
\centering
\includegraphics[width=\textwidth]{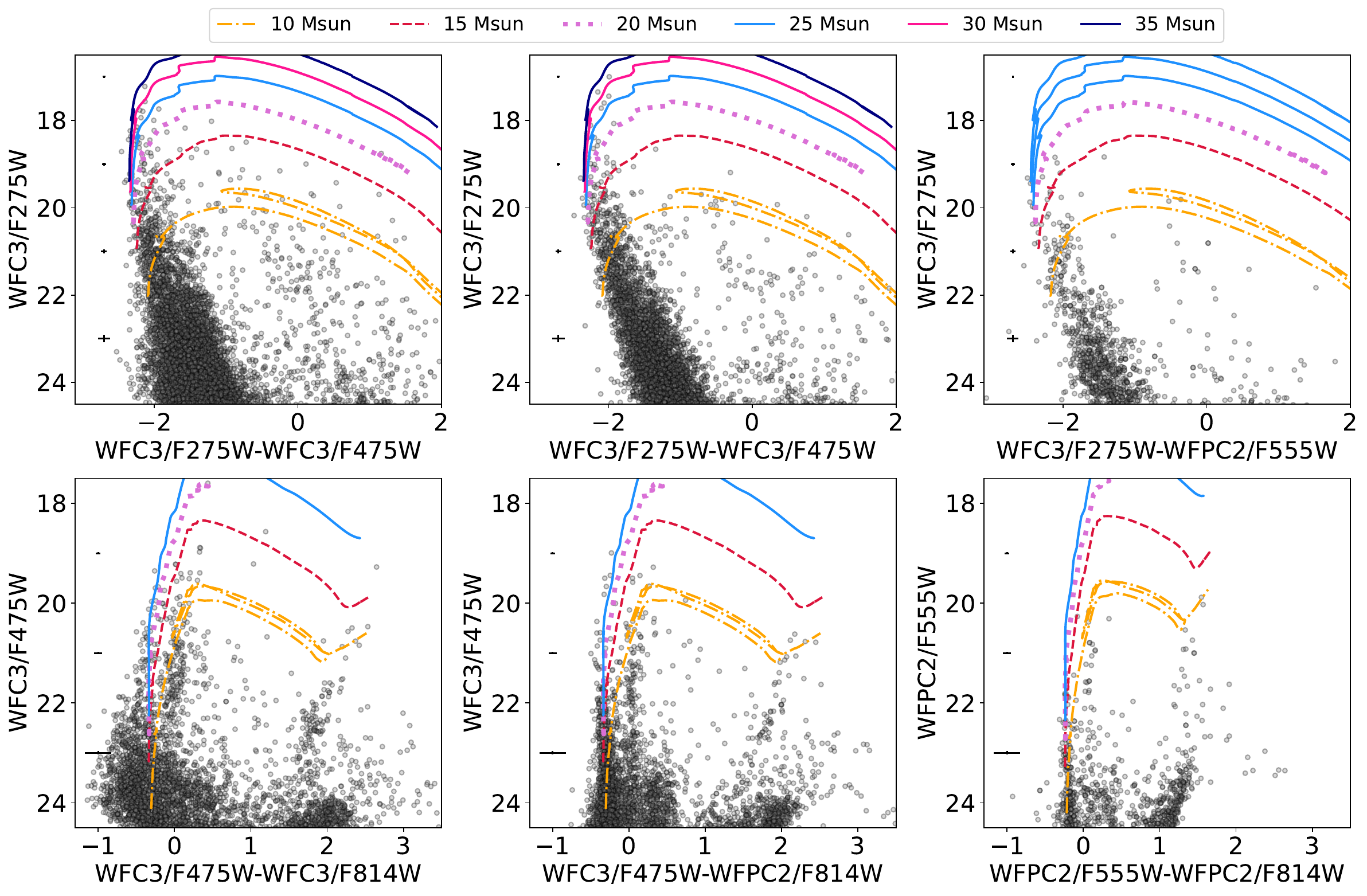} 
 \caption{Select \hst-based UV-optical (top) and optical-only (bottom) CMDs of Sextans~A. We overplot select MIST evolutionary tracks for massive stars, assuming [Fe/H] =$-$1.22, v/vcrit=0.4, and A$_V =0.112$. We indicate the typical errors for each CMD in the left-hand of each subplot.}\label{cmd}
\end{figure*}

Fig.~\ref{cmd} shows select UV and optical CMDs of Sextans A. The CMDs are close to $50\%$ of the completeness limit. We overplot the rotating (v/vcrit=0.4) MESA Isochrones \& Stellar Tracks (MIST) \citep[][]{dotter16,choi16} for reference. We shift the tracks to the J-AGB distance modulus of $\mu =25.71 \pm 0.15$~mag reported by \citet{Lee24}, which is consistent with both the TRGB distance modulus of $\mu=25.76 \pm 0.07$~mag \citep{Lee24} and the Cepheid distance modulus of $\mu=25.61 \pm 0.07$~mag \citep{dolphin03}. We also apply a Milky Way foreground extinction of $A_V=0.122$~mag, consistent with the foreground maps of \citet{schlafly11}.

\section{BEAST fitting} \label{sec:beast}

We fit the panchromatic photometry from the \hst\ broad-band multi-wavelength imaging with the Bayesian Extinction and Stellar Tool \citep[\beast;][]{Gordon16} to infer physical properties of the stars in Sextans~A. In short, the \beast\ uses Bayesian inference to simultaneously model individual stars' stellar and line-of-sight dust properties while considering the observational biases arising from noise and crowding. 

A detailed technical description of the \beast\ can be found in \citet{Gordon16}. The \beast\ has also been used to study properties of nearby low-metallicity dwarf galaxies and both their stellar and dust content \citep[e.g.,][]{choi20, Gull22,Lindberg24, Lindberg25}

Here, we briefly outline the steps taken for this \beast\ fitting. First, a physics model grid is created by mapping a stellar evolution library to a stellar atmospheric model. Initial mass $M_{\rm ini}$, age A, and metallicity $Z$ from the stellar evolution library are mapped to the effective temperature \teff, surface gravity \logg\ and metallicity $Z$ provided by the stellar atmospheric models. We use the PARSEC stellar evolutionary models \citep{bressan12,chen14, Chen_etal_2015}. Furthermore, we adopt a metallicity of $Z=0.06~$Z$_{\odot}$ throughout the paper based on the gas-phase metallicity measurement from \citet{Berg12}. 

The stellar atmosphere models are split into a lower temperature regime (i.e., A-type and cooler), where stars are fit with local thermodynamic equilibrium models from \citet{castelli04}, and a higher temperature regime, where stars are paired with non-local thermodynamic equilibrium (non-LTE) atmospheres from the TLUSTY OSTAR and BSTAR grids (\citealt{lanz03}, \citealt{Lanz07}). We fix the distance modulus for each star to $\mu=25.77$~mag and then apply a dust model that attenuates the intrinsic light. For the dust model, we opt to choose an SMC-like attenuation curve with $R_V = 2.74$ \citep[][]{gordon03} and allow \av\ to vary from a minimum of $A_V=0.00$~mag to a maximum of $A_V=2.0$~mag. 

Second, we split our observations and the AST catalog into three density bins using the spatial density information. Using the resulting ASTs, we create an observational noise model for each resulting dust extinguished stellar model.

Finally, we report the median value (\texttt{$\_$p50}) and the uncertainty computed from the 68\% confidence interval, i.e., $0.5*$(\texttt{$\_$p84}-\texttt{$\_$p16}), of a marginalized 1D posterior probability distribution function (PDF) for each parameter. Throughout the analysis, we report the following parameters: effective temperature (\teff, $\log_{10} (T/[K])$), surface gravity (\logg,  $\log_{10}[\rm cgs]$), extinction (\av, mag), initial mass ($M_{\rm ini}$, \msun), present-day mass ($M_{\rm act}$, \msun), luminosity ($\log(L)$, $\log_{10} (L/[L_{\odot}])$), age ($\log(A)$, $\log_{10} (A/[\rm yr])$) and radius ($R$, $R_{\odot}$). 

As shown in Figure \ref{pos}, not all stars have seven filters (F225W through F160W) coverage; a subset of stars have only 4--5 filters (no \hst\ IR). We split the LUVIT catalog by filter coverage, creating 12 sub-catalogs. These sub-catalogs are all fit independently in the \beast\ with the corresponding ASTs, which will have the same coverage as the data catalog. Whenever possible, we use the seven filter fit for our analysis, then in decreasing order, with a preference for WFC3 F475W and F814W filter coverage over WFPC2 F555W and F814W, to minimize effects from different detectors across all filters used in the SED fitting and to improve spatial resolution. 
We only fit stars covered with at least two UV filters. The various filter combinations which are ultimately used for the analysis (10 combinations) are described in Table \ref{filtercombo} and their spatial distribution is shown in Fig. \ref{pos}. Although we fit a larger set of stars, we report in Table \ref{tab:beast} only the \beast\ results for the main-sequence massive star and OBe star candidates. 

\begin{figure*}[ht!]
\centering
\includegraphics[width=\textwidth]{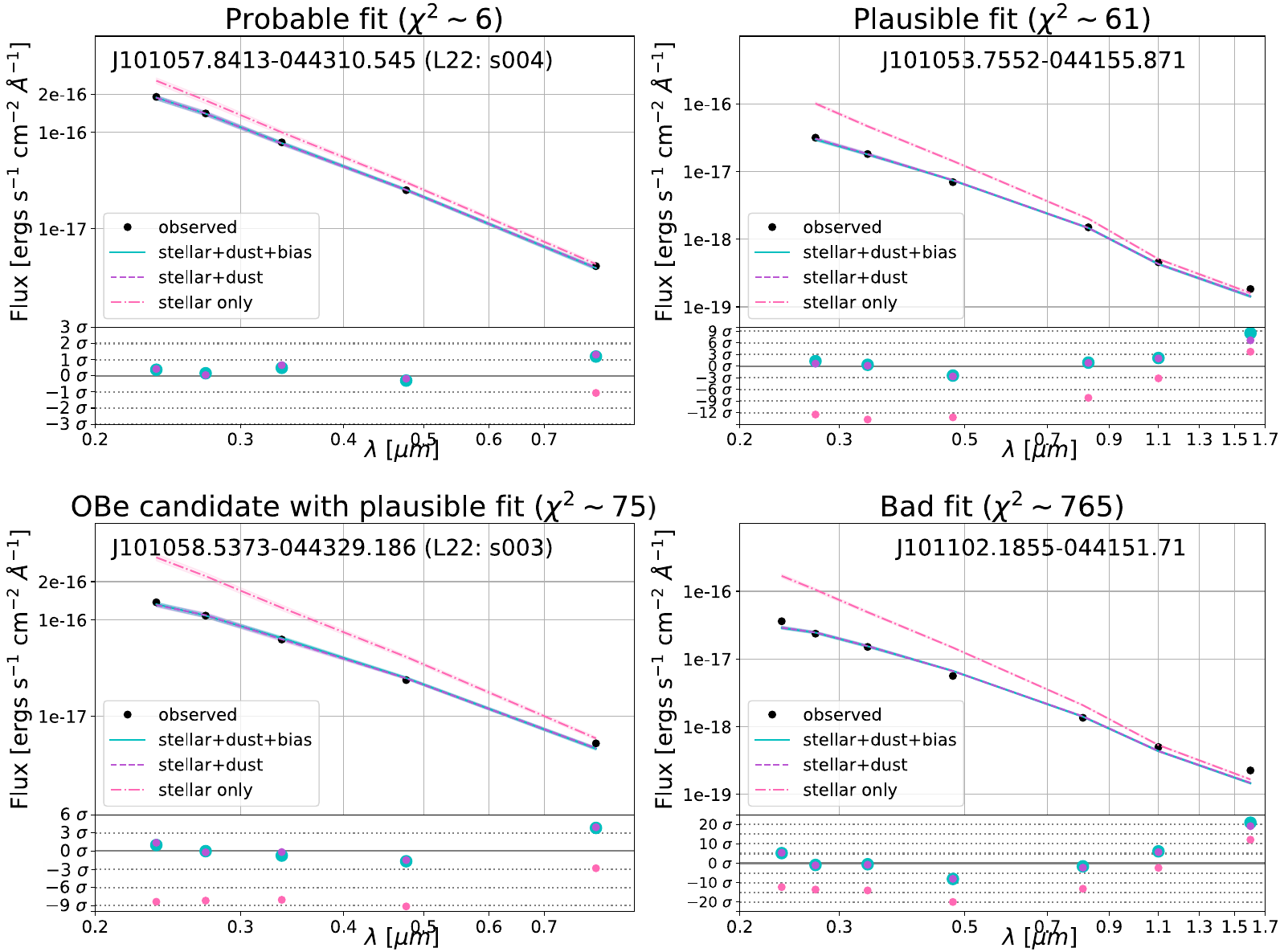} \label{examlpefits}
 \caption{Example SED fits in each category of $\chi^2$ regimes: probable (top left), plausible (top right), plausible marked as OBe candidate (bottom left), and bad (lower right). The data are shown as black points. The median fit stellar (pink), stellar$+$dust (purple), and stellar$+$dust$+$bias (cyan) are overplotted. The sub-panels show the residuals in units of $\sigma$. 
The top left panel shows all residuals of the best model (cyan) within 1.5$\sigma$. The top right panel shows residuals of the best model (cyan) for all UV+optical filters within 3 $\sigma$; only the IR filter \hst/F160W shows a large discrepancy and lies within 9 $\sigma$. The bottom left panel shows residuals of the best model (cyan) within 3 $\sigma$, except for the \hst/F814W filter, where the residual is within 6 $\sigma$. This star is identified as an OBe star candidate, which is likely why \hst/F814W is not well-captured by the underlying model that does not include a disk component. The bottom right panel shows the largest deviations in the residual with the best model (cyan) being within 22 $\sigma$.} 
\end{figure*}

We divide the resulting photometric SED-fitting results into three groups based on $\chi^2$ values to facilitate discussion and comparison. To determine the goodness of fit, we calculate the 50th percentile and 80th percentile of the $\chi^2$ values of the data set. Fits with $\chi^2 <$ 10 (p50) are considered reliable/plausible SED fits, 10 (p50) $ <\chi^2 <$ 100  (p80) are plausible SED fits, and $\chi^2 >$ 100 are poor fits. These $\chi^2$ divisions are consistent with the qualitative assessment of the fits, following other BEAST papers in the literature \citep[e.g.,][]{vandeputte20, Gull22}. Throughout the discussion, we report plausible SED fits and probable SED fits. However, we note that the latter should be considered with caution. We include example fits in Fig. \ref{examlpefits}, we show a probable ($\chi2<10$), plausible ($10<\chi2<100$), and a bad ($\chi2>100$) fit. We also show an example of a plausible ($10<\chi2<100$) fit of an identified OBe star candidate, hence why plausible fits should be considered with caution. Furthermore, we note that the uncertainties are likely underestimated, since they take the models at face value. 

In Fig.~\ref{examlpefits}, the top left panel shows a case of a probable SED fit ($\chi^2 \sim 6$ for the stellar, dust, and crowding bias fit). The blue SED points for the median model are consistent with the observed data within 1-$\sigma$, the reddest filter, \hst/F814W in this case, is within 1.5-$\sigma$. The stars derived \teff and $\log(L)$ are consistent with the Spectral Typing (O5) provided by \citet{lorenzo_new_2022}. However, it may also be due to the uncertainties of dust models at low metallicity.

The right panel of Fig.~\ref{examlpefits} illustrates a plausible fit ($\chi^2 \sim 61$). The residuals show that the median model SED is within $\sim3$-$\sigma$ for the five bluest bands, while the reddest band, \hst/F160W, is consistent within $\sim9$-$\sigma$. The star, however, appears to have quite high extinction (\av $= 0.53$). This may be an artifact of fixing the $R_V$ value in our fitting (See Appendix \ref{sec:fix}).

The bottom left panel of Fig.~\ref{examlpefits} displays an example of a plausible SED fit ($\chi^2 \sim 75$) for star s003, which is an OBe candidate. Compared to the other plausible fit, the deviation from the model occurs already at \hst/F814W, where the models are no longer consistent within $\sim3$-$\sigma$, which is likely due to the disk contribution to the observed SED. 

Lastly, the bottom right panel of Fig.~\ref{examlpefits} displays an example of a bad SED fit ($\chi^2 \sim 765$). In general, the median model values range in agreement from $\sim5$-$\sigma$ for the \hst/F275W, \hst/F336W SED points to $\sim23$-$\sigma$ for the \hst/F160W SED point. At this point, we do not further investigate these types of sources, since it is beyond the scope of this paper, and disregard them in this analysis.

\startlongtable
\begin{deluxetable*}{ccccccccccccc}
\label{tab:beast}
\tablewidth{0pt}
\tablecaption{\label{filtercombo} Filter Combinations}
\tablehead{ 
\colhead{\# Filters} &\colhead{F225W} & \colhead{F275W}&\colhead{F336W} & \colhead{F475W}& \colhead{WFPC2 F555W}&\colhead{WFC3 F814W} &\colhead{WFPC2 F814W} &\colhead{F110W}&\colhead{F160W}&\colhead{N Stars}}
\startdata
7 & \checkmark & \checkmark & \checkmark & \checkmark & \nodata &  \checkmark & \nodata &  \checkmark & \checkmark & 2642\\
6 & \nodata & \checkmark & \checkmark & \checkmark & \nodata &  \checkmark & \nodata &  \checkmark & \checkmark& 3\\
5 & \nodata & \checkmark & \checkmark & \checkmark & \nodata &  \checkmark & \nodata &  \nodata & \nodata & 6456\\
4 & \nodata & \checkmark & \checkmark & \checkmark & \nodata &  \checkmark & \nodata & \nodata & \nodata & 6\\
6 & \nodata & \checkmark & \checkmark & \checkmark & \nodata &  \checkmark & \checkmark &  \checkmark & \checkmark & 4505\\
5 & \checkmark & \checkmark & \checkmark & \checkmark & \nodata &  \checkmark & \checkmark &  \nodata & \nodata & 1\\
4 & \checkmark & \checkmark & \checkmark & \checkmark & \nodata &  \checkmark & \checkmark &  \nodata & \nodata & 1221\\
6 & \nodata & \checkmark & \checkmark & \nodata & \checkmark &  \nodata & \checkmark &  \checkmark & \checkmark& 53\\
5 &  \checkmark & \checkmark & \checkmark & \nodata & \checkmark &  \nodata & \checkmark &  \nodata & \nodata& 1\\
4 & \nodata & \checkmark & \checkmark & \nodata & \checkmark &  \nodata & \checkmark &  \nodata & \nodata & 106\\
\enddata
\tablecomments{Summary of various filter combinations of the Sextans A data used in this work. When overlap occurs in the data, we choose HST/WFC3 over the HST/WCPC2.}
\end{deluxetable*}

\section{Massive stars in Sextans A} \label{sec:mstar}

 We find 500 massive stars with $M_{\rm ini,\,\rm p50}>8$\msun\ and log$(g)_{\rm p50}>3.7$ in our panchromatic analysis with $\chi^2$ $<10$ and 867 in this mass range with $\chi^2$ $<100$. We note that 13773 stars in total (no cuts) are in the $\chi^2$ $<100$ regime. No other cuts have been applied thus far. For this part of the discussion, we will focus on main-sequence stars, as massive stars beyond the main sequence are not always well captured by current evolutionary models. These large uncertainties are driven by physical phenomena like binarity \citep[e.g.,][]{Langer12}, rotation  \citep[e.g.,][]{Szecsi15}, or overshooting \citep[e.g.,][]{Bressan15}, which can heavily impact their evolution at low metallicity.
 
\begin{figure*}[ht!]
\centering
\includegraphics[width=\textwidth]{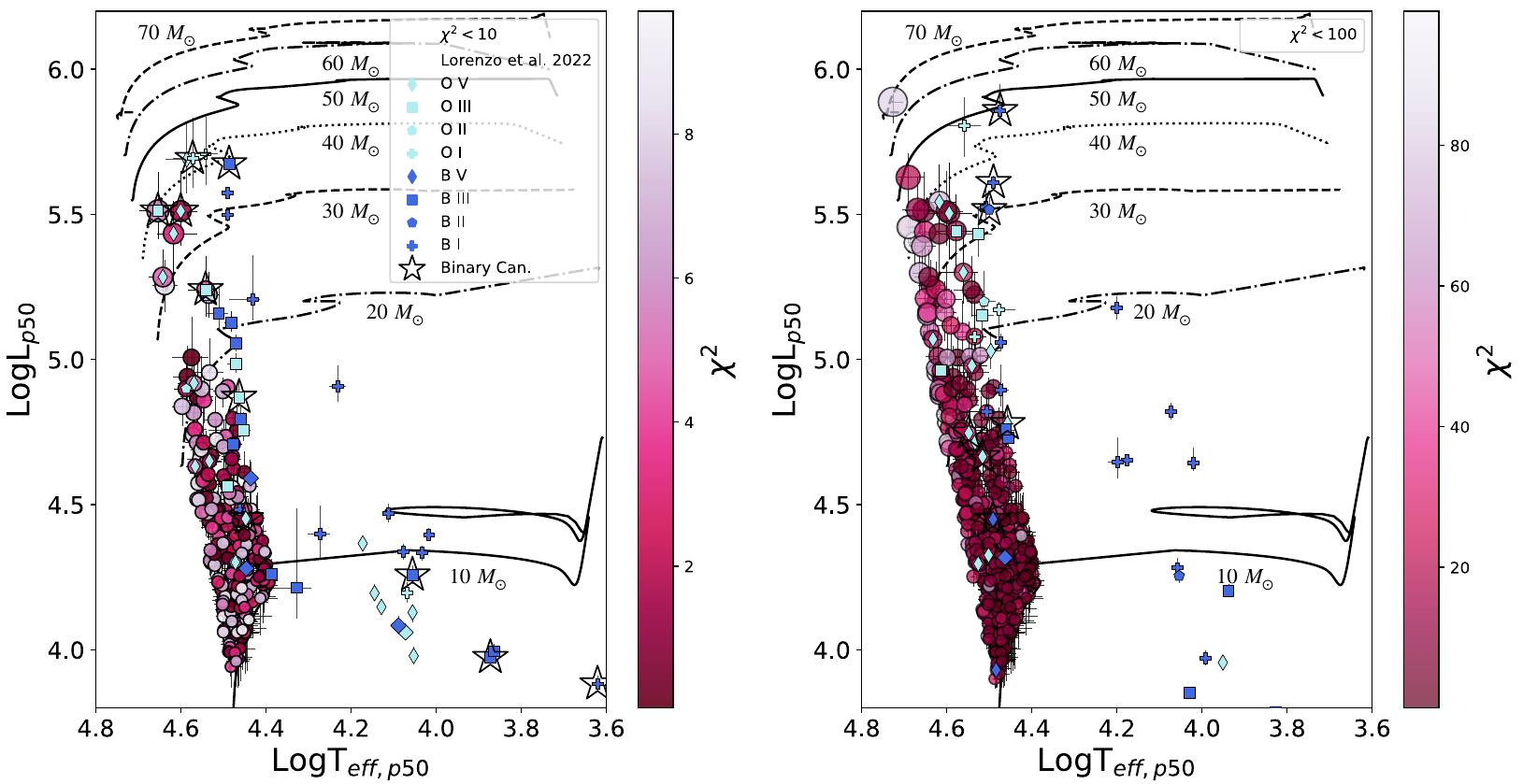} 
 \caption{Hertzsprung-Russell diagram of the main-sequence massive stars in Sextans A ($M_{\rm ini,\,\rm p50}>8$\msun\ and log$(g)_{\rm p50}>3.7$). We overplot Parsec evolutionary tracks. The left panel is for stars where the \beast\ yields $\chi^2<10$, whereas the right panel represents stars with $10< \chi^2<100$. The stars are color-coded by their $\chi^2$ values, and the size of the stars is scaled to their mass. Additionally, we overplot the stars in the \citet{Lorenzo22} sample, color-coded by spectral and luminosity type. We mark potential binary candidates as stars. The stellar parameters used to plot the \citet{Lorenzo22} sample are those we derive for the stars using the \beast. There is a subset of stars that do not coincide with our main-sequence sample, which is why they do not appear to have a counter match with our \beast\ sample (e.g., giants, supergiants). Several stars that appear at odd locations for their spectral type are among the faintest stars in the sample, and the low SNR may have complicated the spectral typing.} \label{HR}
\end{figure*}

 Figure \ref{HR} shows the sample of massive main-sequence stars. We find 5 above 30 \msun, 7 between 20 \msun\ and 30 \msun, and 232 between 10 \msun\ and 20 \msun for stars with with $\chi^2$ $<10$. For stars with $10<$ $\chi^2$ $<100$, we find 13 above 30 \msun, 43 between 20 \msun\ and 30 \msun, and 179 between 10 \msun\ and 20 \msun. Lastly, we have one star with $\chi^2$ $<100$ between 60 \msun\ and 70 \msun\ and four between 40 \msun\ and 50 \msun. Table \ref{mostmassive} summarizes the masses discussed in this paragraph, and Table \ref{allstars} reports all the masses and selected fitting parameters.

\startlongtable
\begin{deluxetable}{ccc}
\tablecaption{Summary of massive main-sequence star in Sextans A with log$(g)_{\rm p50}>3.7$}\label{mostmassive}
\tablehead{\colhead{Mass Range} & \colhead{Num. of stars} & \colhead{Num. of stars}\\
\colhead{} & \colhead{$\chi^2<10$} & \colhead{$10<\chi^2<100$}
}
\startdata
$60$ \msun $- 70$ \msun & 0 & 1\\
$40$ \msun $- 50$ \msun & 0 & 4 \\
$30$ \msun $- 40$ \msun & 5 & 8 \\
$20$ \msun $- 30$ \msun & 7 & 43 \\
$10$ \msun $- 20$ \msun & 232 & 179 \\
$8$ \msun $- 10$ \msun & 256 & 132 \\
$>8$ \msun & 500  & 367 \\
\enddata
\end{deluxetable}

No spectra exist for 454 of the massive stars with $\chi^2$ $<10$  or 829 for massive stars with $\chi^2$ $<10$. Above 25 \msun\ 4 out of 6 stars ($\chi^2$ $<10$ ) and 9 out of 31 stars  ($\chi^2$ $<100$)  currently have low resolution spectra \citep{Lorenzo22}, and one star has medium resolution spectra \citep{telford21}. In Figure \ref{HR}, we plot the massive MS in our sample and show the massive star candidates identified by \citet{Lorenzo22}, including giants and supergiants, which are not all MS stars. We note that on the upper end of the HR diagram (log$(L)>4.5$), most stars in both panels are located in reasonable areas based on their spectral type. Exceptions include stars marked as binary candidates (see Appendix \ref{sec:singletracks}), which may have misleading stellar parameters despite their plausible SED fits. In particular, the B III type star appearing to the right of the supergiants in the left panel may appear more luminous in the \beast\ fit due to extra light contamination from the secondary. The lower end of the HR diagram shows more peculiar placements of stars for their spectral type. We note that a subset of these stars, particularly the O V star residing to the right of the main-sequence, are from spectra that are reported with quite low S/N, which may have led to spectral misclassifications, especially for stars in crowded regions. Lastly, the HR diagram shows four stars classified as supergiants residing on the MS. In \citet{lorenzo_new_2022}, these stars (s053, s073, s075, s116) are all described as having contaminated spectra, either a blend or a double-lined spectroscopic binary (SB2), which may make the luminosity classification harder. We will further compare with the literature in Section \ref{comp}.

%\startlongtable
\begin{deluxetable*}{cccccccccccccccccccccccccccc}
\rotate
\tablecaption{\beast\ fit for the massive stars in Sextans A}\label{allstars}
\tablehead{\colhead{LUVIT ID} & \colhead{$M_{\rm ini,\,\rm p50}$} & \colhead{$M_{\rm act,\,\rm p50}$} & \colhead{$\log(L)_{\rm p50}$} & \colhead{$\log(T)_{\rm p50}$} &  \colhead{\logg$_{\rm p50}$} & \colhead{$\log(A)_{\rm p50}$} & \colhead{radius$_{\rm p50}$} & \colhead{$A_{V,\,\rm p50}$} & \colhead{$\chi^2_{min}$} & \colhead{Filters} & \colhead{Literature}\\
\colhead{} & \colhead{\msun} & \colhead{\msun} & \colhead{dex [1/$L_{\odot}$]} & \colhead{dex [K]} &  \colhead{dex [cgs]} & \colhead{dex [yr]} & \colhead{R$_{\odot}$} & \colhead{dex} & \colhead{} & \colhead{}}
\startdata
\hline
&&&&& $\chi^2<10$ &&&\\
\hline
J101057.84$-$044310.55 & 39.32$_{-5.36}^{+10.42}$ & 39.01$_{-4.59}^{+10.43}$ & 5.51$_{-0.1}^{+0.14}$ & 4.65$_{-0.03}^{+0.04}$ & 4.09$_{-0.1}^{+0.14}$ & 6.5$_{-0.28}^{+0.11}$ & 9.32$_{-0.43}^{+0.39}$ & 0.14$_{-0.02}^{+0.02}$ & 5.63 & 5 & L22: s004 \\
J101106.43$-$044237.42 & 35.55$^{+8.69}_{-5.15}$ & 35.38$^{+7.67}_{-5.51}$ & 5.51$^{+0.14}_{-0.12}$ & 4.6$^{+0.05}_{-0.04}$ & 3.83$^{+0.14}_{-0.1}$ & 6.65$^{+0.09}_{-0.12}$ & 11.91$^{+0.57}_{-0.77}$ & 0.16$^{+0.03}_{-0.02}$ & 1.74 & 5 & L22: s007 \\
J101105.16$-$044240.84 & 33.64$^{+6.89}_{-6.9}$ & 34.12$^{+6.95}_{-7.32}$ & 5.43$^{+0.11}_{-0.17}$ & 4.62$^{+0.04}_{-0.06}$ & 3.95$^{+0.11}_{-0.17}$ & 6.63$^{+0.15}_{-0.13}$ & 10.14$^{+0.65}_{-0.49}$ & 0.11$^{+0.03}_{-0.03}$ & 3.27 & 5 & L22: s023 \\
J101104.24$-$044238.34 & 32.05$^{+6.12}_{-3.6}$ & 31.41$^{+6.0}_{-2.8}$ & 5.28$^{+0.09}_{-0.07}$ & 4.64$^{+0.03}_{-0.02}$ & 4.18$^{+0.1}_{-0.07}$ & 6.51$^{+0.1}_{-0.28}$ & 7.53$^{+0.27}_{-0.24}$ & 0.06$^{+0.02}_{-0.02}$ & 4.99 & 5 & L22: s129 \\
J101058.33$-$044040.14 & 30.97$^{+3.95}_{-4.19}$ & 30.56$^{+4.93}_{-3.73}$ & 5.26$^{+0.09}_{-0.09}$ & 4.64$^{+0.03}_{-0.03}$ & 4.17$^{+0.09}_{-0.08}$ & 6.54$^{+0.13}_{-0.18}$ & 7.47$^{+0.27}_{-0.25}$ & 0.62$^{+0.02}_{-0.02}$ & 8.56 & 6 & \nodata \\
J101058.15$-$044318.71 & 25.27$^{+5.07}_{-2.21}$ & 25.08$^{+4.87}_{-2.29}$ & 5.24$^{+0.12}_{-0.09}$ & 4.54$^{+0.04}_{-0.02}$ & 3.73$^{+0.12}_{-0.08}$ & 6.82$^{+0.06}_{-0.1}$ & 11.26$^{+0.56}_{-0.64}$ & 0.13$^{+0.03}_{-0.03}$ & 4.4 & 5 & L22: s029, T24: S3 \\
J101053.76$-$044113.44 & 24.71$^{+3.35}_{-2.04}$ & 24.49$^{+3.03}_{-2.44}$ & 5.22$^{+0.08}_{-0.08}$ & 4.53$^{+0.03}_{-0.03}$ & 3.71$^{+0.08}_{-0.09}$ & 6.83$^{+0.06}_{-0.06}$ & 11.57$^{+0.56}_{-0.44}$ & 0.06$^{+0.03}_{-0.02}$ & 6.55 & 6 & L22: s014 \\
J101056.99$-$044019.92 & 22.46$^{+4.8}_{-3.26}$ & 22.39$^{+4.65}_{-3.22}$ & 5.01$^{+0.14}_{-0.12}$ & 4.58$^{+0.05}_{-0.04}$ & 4.03$^{+0.13}_{-0.11}$ & 6.79$^{+0.13}_{-0.19}$ & 7.52$^{+0.44}_{-0.35}$ & 0.53$^{+0.03}_{-0.03}$ & 0.08 & 4 & \nodata \\
J101054.42$-$044048.20 & 21.88$^{+3.76}_{-2.95}$ & 21.71$^{+3.84}_{-2.92}$ & 4.94$^{+0.1}_{-0.1}$ & 4.58$^{+0.03}_{-0.04}$ & 4.13$^{+0.1}_{-0.1}$ & 6.74$^{+0.14}_{-0.19}$ & 6.65$^{+0.31}_{-0.27}$ & 0.14$^{+0.03}_{-0.03}$ & 0.03 & 4 & \nodata \\
J101106.47$-$044139.06 & 21.44$^{+2.99}_{-2.56}$ & 21.29$^{+3.22}_{-2.45}$ & 4.9$^{+0.09}_{-0.09}$ & 4.59$^{+0.03}_{-0.03}$ & 4.18$^{+0.09}_{-0.09}$ & 6.71$^{+0.13}_{-0.19}$ & 6.23$^{+0.25}_{-0.23}$ & 0.17$^{+0.02}_{-0.02}$ & 1.57 & 5 & L22: s017 \\
J101059.59$-$044024.36 & 21.39$^{+2.71}_{-2.12}$ & 21.22$^{+3.04}_{-2.24}$ & 4.84$^{+0.09}_{-0.07}$ & 4.6$^{+0.02}_{-0.02}$ & 4.26$^{+0.08}_{-0.07}$ & 6.61$^{+0.15}_{-0.28}$ & 5.64$^{+0.19}_{-0.17}$ & 0.68$^{+0.02}_{-0.02}$ & 7.49 & 6 & \nodata \\
J101106.61$-$044213.36 & 20.82$^{+3.53}_{-2.2}$ & 20.56$^{+3.75}_{-2.04}$ & 4.92$^{+0.12}_{-0.08}$ & 4.57$^{+0.04}_{-0.03}$ & 4.06$^{+0.11}_{-0.08}$ & 6.81$^{+0.1}_{-0.16}$ & 6.99$^{+0.26}_{-0.29}$ & 0.07$^{+0.02}_{-0.02}$ & 5.55 & 5 & L22: s008 \\
\nodata & \nodata & \nodata & \nodata & \nodata & \nodata & \nodata & \nodata & \nodata & \nodata & \nodata \\
\enddata
\tablecomments{We show here a subset of the catalog with \beast\ fits, which yield $M_{\rm ini,\,\rm p50}>8~M_{\odot}$. \cite{lorenzo_new_2022} is denoted as L22, and \cite{telford_observations_2024} is denoted as T24. The full \beast\ fit catalog is available as machine-readable tables (mrt).}
\end{deluxetable*}

\section{Comparison to Stellar Spectra in Literature}  \label{comp}
\subsection{Spectral Typing and Panchromatic Photometry}  \label{sec:Spt}

\begin{figure*}[ht!]
\centering
\includegraphics[width=\textwidth]{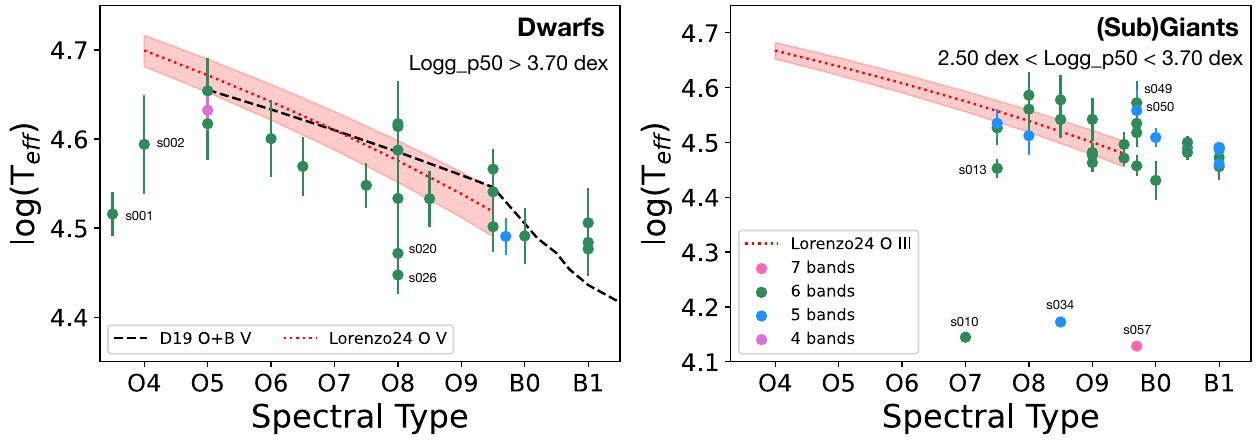} 
 \caption{ \teff\ is plotted vs.\ spectral type for early-type (O-B3) stars in our sample. The \teff\ values are taken from our \beast\ fit, while the Spectral Types are the types reported in \citep{Lorenzo22}. We classify stars with \logg$_{\rm p50} >3.70 $ as dwarfs (left), and $2.50 <$ \logg$_{\rm p50} < 3.70 $ as giants. We note that these classifications are based solely on photometry and must be confirmed by spectra. The stars are color-coded by the number of filters they were observed in. We compare our stars ($\chi^2 \leq 100$) with the theoretical \teff\ vs.\ Spectral Type relation derived by \citet{Lorenzo25} as well as the \teff\ scale derived by \citet{Dufton19} (dashed line, D19). The stars follow the relationship within error bars, except for four dwarfs and six (sub)giants.} \label{specteff} 
\end{figure*}

Spectral typing can be a useful qualitative tool for describing massive stars for which only low-resolution spectra are available. As higher-resolution spectra become available, stellar parameters are updated, and spectral types are corrected. However, even low-resolution spectra can be observationally expensive for metal-poor massive stars. A more cost-effective alternative for this initial spectral characterization is panchromatic photometry \citep{Gull22}. For most massive stars appearing to be non-binary, the photometric stellar parameters and spectroscopic stellar parameters appear consistent, and outliers can routinely be flagged in the panchromatic photometry. A subset will not appear as suspicious in the panchromatic photometry (See Appendix \ref{sec:singletracks}). However, the vetting sample is currently still limited to a dozen stars.\\

We further test this method by comparing a subset of our derived panchromatic stellar parameters to the spectral typing of \citet{Lorenzo22}. We focus on the early-type (O-B3) stars. First, we use the commonly used spectral type-effective temperature relationship. We divide the stars that yield $\chi^2$ $<100$ into two luminosity classes (dwarfs and giants) using the derived \logg\ values. Any star with \logg\ $> 3.70$ is designated a dwarf, while stars between $3.70>$ \logg\ $> 2.50$ are designated as a giant. We do not include a comparison of the supergiants. In Fig. \ref{specteff}, we present 22 dwarfs and 32 (sub)giants in our sample, with spectral types reported in \citet{Lorenzo22}. In addition to our data, we include commonly used observational relations between spectral type and temperature reported by \citet{Dufton19} for the SMC and the theoretical prediction of the relationship reported by \citep{Lorenzo25} for sub-SMC metallicity.

\subsubsection{Dwarfs} 
Of the 22 dwarfs, we derive slightly lower temperatures for a given spectral type. For most of them, the error bars fall within the uncertainties derived by \citet{Lorenzo25}. We find four clear outliers showing lower temperatures than expected for the reported spectral type in \citet{Lorenzo22}. These stars are s001, s002, s020 and s026. We note that for s001 and s002 we derive $\chi^2$ values of 24.5 and 18.7, respectively. This may be unsurprising as \citet{Lorenzo22} flagged s001 as a potential double-lined spectroscopic binary (SB2) system and noted strong He II absorption and nebular emission. Both stars may have companions undetected by the \beast, especially if the companion is hotter and less massive, since it could easily hide in the photometry. s020 is classified as either a double-lined spectroscopic binary (SB2) or a blend, and s026 is considered a potential blend. The \beast\ fits yield $\chi^2$ $<10$. The disagreement here could be explained by the lack of resolution in the spectra compared to the high spatial resolution in the photometry. 

\subsubsection{(Sub)Giants}
We identify 6 stars among the 32 (sub)giants that significantly deviate from the theoretical relation. The temperatures derived for the stars s010, s013, s034, and s057 are too cool, while s049 and s050 are too hot. s010 has very low SNR and resolution, while s034 also has low SNR and is classified as an O-type star despite the absence of He II 4686. Neither the s013 spectra nor the \beast\ fit suggest any peculiarities. s050 appears to be a potential binary. We note that s050 and s049 are classified as supergiants in \citet{Lorenzo22}. It has previously been suggested that stellar atmosphere models do not adequately capture supergiants \citep{mcquinn12a,Gull22}, so this may be a plausible explanation for the classification as a (sub)giant by the photometry alone. Overall, the \beast\ recovers temperatures that align with the spectral typing from low-resolution spectra.

\subsection{Luminosity Classes and CMD Analysis}
\begin{figure}[ht!]
\centering
\vspace{-0.1cm}
\includegraphics[width=\columnwidth]{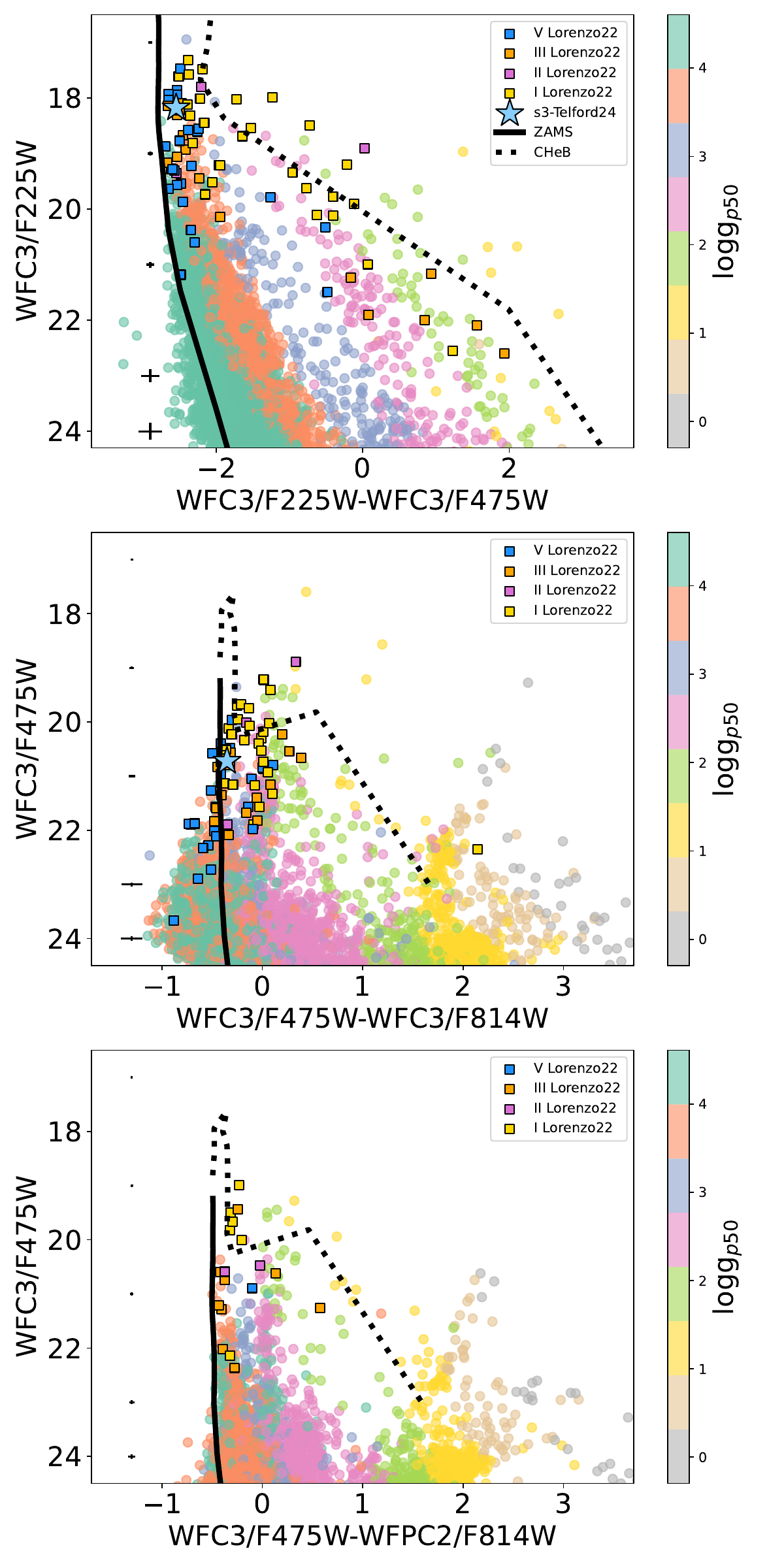} 
\caption{\small We show NUV-Optical (top) and Optical-only (middle and bottom) CMDs of Sextans~A. The stars are color-coded by the \logg$_{\rm p50}$ values obtained by the \beast\ fit. Furthermore, we overplot the stars observed in \citet{telford21} (star) and \citet{Lorenzo22} (squares), the latter of which are color-coded by their luminosity class based on the R$\sim 1000$ spectra in \citet{Lorenzo22}. In the optical, distinguishing between surface gravity and luminosity class is challenging; however, the NUV-Optical CMD paints a clearer picture. Most stars are placed on the CMD as expected based on their luminosity class, with a couple of outliers.} \label{cmdspt}
\end{figure}

We also explore the luminosity classes reported in \citet{Lorenzo22} in relation to their positions in the CMD and the derived surface gravities from the \beast\ in Figure \ref{cmdspt}. We also include S3 (s029) analyzed in \citet{garcia19a,telford24,Furey25}, which we will discuss in Section \ref{gracestar}. In the optical, the various surface gravities are nearly impossible to distinguish due to blending, especially in the high mass regime. However, in the NUV-optical CMD, we see a clearer trend in surface gravity. Although the surface gravity values depend on the underlying stellar atmosphere models, the trend across the CMD will be accurate for most stars.

We find that the majority of dwarfs (denoted as V) identified by \citet{Lorenzo22} fall in the $4.5$ dex $>$ \logg$_{\rm p50}$ $>3.7$ dex range of the CMD. However, three stars classified as dwarfs appear redward of the main-sequence. We note that in \citet{Lorenzo22}, it is remarked that one was observed with the 1.5 slit (very low resolution, R$<1000$), one seemed to be a blend (potential binary), while the last star has no special remarks but is quite faint. 

Furthermore, we find that three (s003, s024, s043) stars marked as dwarfs and on the main-sequence in the $4.5$ dex $>$ \logg$_{\rm p50}$ $>3.7$ dex region in the NUV-Optical CMD appear in the $3.0$ dex $>$ \logg$_{\rm p50}$ $>2$ dex region in the optical only CMD. In \citet{Gull22}, this phenomenon was described as typical for Be-type stars. The spectra of s003 show H$\beta$ and H$\gamma$ emission, as well as a relatively shallow He I line. This combination indicates a potential high \vsini, a spectral feature typically associated with Oe/Be stars. We discuss these further in Section \ref{sec:bestars}.
Some stars identified as supergiants (I) appear blueward of the core Helium burning sequence (cHeB). The cHeB may not be well calibrated, as underlying assumptions in the evolutionary tracks can significantly impact its location. We only comment on the faintest and bluest five sources, which \citet{Lorenzo22} comments on potential contamination by either a nearby star or nebula that may have impacted the luminosity class determination.

\subsection{S3 -- s029}\label{gracestar}
S3 is one of the most studied stars at sub-$10\%$ \zsun massive stars. The star was first observed with ($R\sim 1000$) as part of the \citet{garcia19a} study of hot stars in Sextans A. Medium-resolution spectra were obtained at R$\sim 6000$ resolution in the UV as part of \citet{telford21}. It was also observed at low resolution (R$\sim 1000$) in \citet{Lorenzo22}. Medium-resolution spectra were obtained at R$\sim 4000$ in the optical as part of \citet{telford24, Furey25}. A challenging problem for the star has been that the photometrically derived parameters disagree with the spectroscopically derived parameters. In \citet{garcia19a}, the star was identified as an O9 V star, although they note that the derived luminosity seemed closer to a III type star. In \citet{Lorenzo22}, the star was classified as an 08.5 III star and denoted as an SB1 candidate. \citet{telford24}, using higher resolution optical and UV spectra, could not confirm this claim, and reclassified the star as an O9 V star. 

The main parameter divergence lies in the surface gravity, and therefore in the masses. In \citet{telford24}, the spectroscopic surface gravity is 3.9$\pm 0.1$, which suggests a spectroscopic mass of 49 $^{+30}_{-19}$ \msun\ while in \citet{Furey25} it was derived to be 4.18$^{+0.14}_{-0.08}$, which suggests a spectroscopic mass of 77.8 $^{+25}_{-12}$ \msun. In contrast, the photometric 3.5$\pm 0.1$ reported by \citet{telford24} yielded a photometric mass of 21 \msun. Our \beast\ fit, which yields $\chi^2<10$, suggest a surface gravity of 3.73$^{+0.12}_{-0.08}$ and mass of 25.1$^{+4.9}_{-2.3}$ \msun. Surprisingly, we find consistent luminosity values throughout with $\log(L)= 5.1$ (photometry) \citet{telford21}, $\log(L)= 5.15$ (spectra) \citet{telford24}, $\log(L) = 5.18^{+0.05}_{-0.04}$ \citet{Furey25} and $\log(L) = 5.24^{+0.12}_{-0.09}$ for our \beast\ fit. Current evolutionary tracks suggest that the masses derived by \citet{Furey25} and \citet{telford24} are too high for the luminosity and temperature derived, making the star under-luminous for its mass, as the authors point out. The tension arises from differences in surface gravity between spectroscopy and photometry, as the spectroscopic mass strongly depends on surface gravity. We note that the photometric surface gravity is model-dependent. \\

While we cannot comment further on the spectroscopy, higher-resolution or multi-epoch spectra may shed more light on this disagreement.  We can further investigate the photometry. In the top two panels of Figure \ref{cmdspt}, we show the position of S3--s029. The NUV-optical CMD falls within the $4.00<$\logg$_{\rm p50} < 3.5$ range of surface gravity values. In the optical, it does indeed appear closer to the main-sequence, but as explained in Section \ref{sec:Spt}, $4.50<$ \logg$_{\rm p50} < 3.0$ overlap in this region of the CMD. Although S3 behaves like a standard single star, we remain cautious, as binaries and binary products can easily hide in photometry. \\

We perform one last test, integrating the light from all the HST sources within the KCWI PSF ($\sim$ 1 arcsec), and re-run the \beast. This test aims to rule out significant contributions to the flux from nearby sources, which would perhaps lead to an increased spectroscopic mass. We find a slightly higher, but still acceptable, $\chi^2$ value and a slightly higher mass ($\sim +1-2$ \msun). Even so, the remaining parameters are relatively consistent with the previous measurement. This is because S3 is a relatively isolated source. It is associated with two other massive stars, which appear to be at least 3 arcseconds away. The stars closest to S3 appear otherwise faint, so S3 still dominates the flux compared to its neighboring sources even if it were to be unresolved in the KCWI spectra. It is unclear at this point whether S3 is just a peculiar source (a binary or a binary product) or whether the single-star models are the culprits and need to be adjusted at low metallicity. The former appears more likely, since the expectation is that massive stars are in binaries or experience binary interactions. Photometrically, monitoring the source with time-series photometry would further constrain the true nature of this star. Fortunately, the Vera Rubin Observatory Legacy Survey of Space and Time (LSST) will map out the southern sky every two weeks over the next ten years. S3 is bright enough to be reliably observed by LSST, so we should expect to be able to monitor S3 for any brightness changes. Spectroscopically, we require both multi-epoch spectra to monitor for a companion and higher-resolution and SNR spectra to measure the surface gravity with higher precision.

\section{OB\lowercase{e} stars fraction and candidates} \label{sec:bestars}
OBe stars are rapidly rotating main-sequence (MS) O/B-type stars surrounded by a disk, which gives rise to emission (e) and IR excess due to bound-free and free-free emission from the ionized gas. OBe stars are cornerstones in the evolution of massive stars and may also play a role in the formation of compact binaries \citep[e.g.,][]{marchant_evolution_2023,Rivinius26}. Yet, their formation is still debated (spin-up through mass transfer vs. high initial rotational velocity) \citep{Pols91,Granada13,rivinius_classical_2013,Hastings20a,Hastings20,Rivinius26}. 

\begin{figure*}[ht!]
\includegraphics[width=\textwidth]{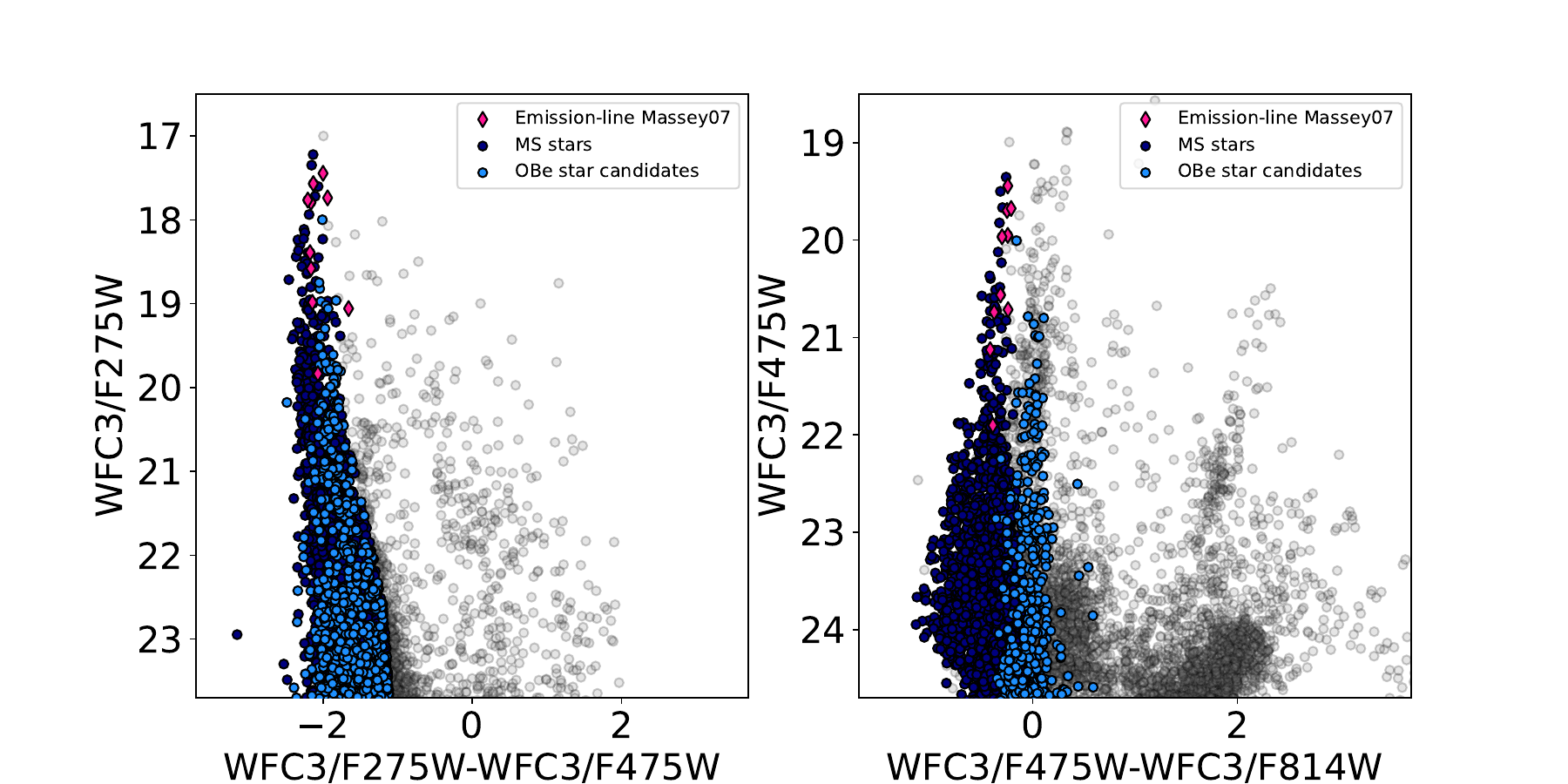} 
\caption{UV-optical (left) and optical (right)  CMDs of Sextans~A. Stars that fall on the main sequence in the UV-optical CMD are marked in navy blue. We plot emission-line stars identified by \citet{Massey07} in pink. We additionally plot photometric OBe star candidates in light blue; these are identified as stars that fall on the UV-optical main sequence but fall redward of the main-sequence in the optical CMD or have F814W excess relative to the \beast\ fits. These classifications imply an OBe fraction of $\sim 15\%$ in Sextans A for stars with $F275W>22.9$, a  OBe fraction of $\sim 23 \%$ for stars $F275W>21$ and a OBe fraction of $\sim 17\%$ for stars $F275W>19.5$.} \label{excess}
\end{figure*}

\subsection{Photometry of OBe stars}
 OBe stars can be identified photometrically through their IR-excess \citep[e.g.,][]{Dougherty94,rivinius_classical_2013} or H$\alpha$ emission \citep[e.g.,]{Kogure07,rivinius_classical_2013}. \citet{Gull22} showed that in CMDs created using \hst\ photometry, OBe star candidates fall on the main-sequence in the UV-optical CMD (\hst/F275W, \hst/F475W), but fall redward of it, near the cHeB, in the optical CMD (\hst/F475W, \hst/F814W). We repeat this analysis for the stars in Sextans A. Using evolutionary tracks, we identify the edge of the main sequence in the UV. We then use two different methods to identify potential OBe stars. First, we identify stars that fall redwards of the MS's edge in the optical CMD. Second, we use the \beast\ simulated photometry. Specifically, we compare the observed F814W magnitude to the simulated F814W magnitude and identify stars that show an excess of more than 3$\sigma$ in the observed magnitude relative to the predicted magnitude from the \beast.

In Figure \ref{excess}, we show the results of applying these cuts and observe a behavior similar to that of Leo A: a subset of MS stars in the UV appears to pollute the BHeB sequence. 

NIR excess is not the only signature of OBe stars in photometry. They can also be identified using H$\alpha$ imaging. Hence, we plot the emission-line stars identified through shallow ground-based H$\alpha$ imaging by \citet{Massey07}. To identify emission-line stars, they first apply a 20th mag cut-off on the H$\alpha$ magnitudes. Second, reject stars with H$\alpha$-[S II]$\geq −0.15$. Third, they use [O III]-
$(V + B)/2  \geq −0.75$ to attempt to reject nebular emission. Lastly, they select for intrinsically blue objects by imposing $(U − B) − 0.72~(B − V ) \geq −0.4$.

We note that in Fig. \ref{excess}, none of the OBe candidates identified through our IR-excess process appear in their sample. Most likely this is due to their H$\alpha$ magnitude and the crowding issue (aperture $\sim1.35^"$). On the other hand, the position in the CMD of their emission-line candidates does not seem to suggest a large IR-excess in the \hst/F814W filter. We consider a few possible reasons for this.

First, Be stars are know to ``turn-on/-off'' their disks on month-to-year timescales \citep{Hubert98,Porter03,deWit06,McSwain08,Peters20, Figueiredo25,Rivinius26}. It is possible that at the time of the H$\alpha$ observations, a disk was present and we would also have observed IR-excess, but at the time of the \hst/F814W observations, the disk was absent, and we would also not have observed H$\alpha$.  While it is unlikely that this is the case for all emission-line candidates of \citep{Massey07}, it may explain some, since this phenomenon affects $\sim 22\%$~of all OBe stars per year as suggested in previous studies \citep{Peters20}.

Second, for the brighter candidates in the sample (\hst/F475W $<20$~mag), it is possible that the flux at \hst/F814W is still dominated by the stellar SED, so much so that we do not detect an IR excess from the disk in the filter. Existing studies of the OBe phenomena are focused on higher metallicity regimes. However, at low metallicity, more luminous Oe stars can be formed more easily \citep{rivinius_classical_2013}, and it is possible that some Oe stars do not show red/IR excess in \hst/F814W. This has been observed in the SMC, where some bright OBe stars still appear on the MS in the optical CMD \citep{Bodensteiner21,Bodensteiner25}. 

Lastly, the extent and slope of the IR excess depend on disk inclination \citep{Carciofi06}. It is thus possible that the emission-line stars identified by \citet{Massey07} are viewed at inclinations which produce little IR excess in \hst/F814W.

With these caveats in mind, we consider both photometric and spectroscopic OBe star candidates in calculating OBe star fractions.  Using reliable $\chi^2$ stars, we determine the cutoff for $\sim$8 \msun, $\sim$15 \msun\ and $\sim$20 \msun\ in F275W magnitude. Using these cutoffs, we find that above $\sim$ 8 \msun\ the lower limit of the OBe fraction is $15\%$ (counting identified emission-line stars), above $\sim$ 15 \msun\ the lower limit of the OBe fraction is  $23\%$ and above $\sim$ 20 \msun\ the lower limit of the OBe fraction is $17\%$. This increase in OBe fraction across mass and metallicity follows similar findings to those of \citet{Peters20}. We also find relatively good agreement with the $27 \pm 3\%$ for Sextans~A derived in \citet{Schootemeijer22}, given that we present a lower limit.

Eleven stars have derived \beast\ masses between $41-81$ \msun\ and fall within our OBe star candidate category. Seven stars show high values of $\chi^2$ but constitute half of the brightest OBe candidates. It is not entirely clear what the actual mass of these systems is, since the high $\chi^2$ values indicate that the \beast\ attempted to increase $Av$ and luminosity to match the observed SED. The UV CMD suggests they are more likely in the $20-40$ \msun\ range; but if we assume they are binary products or experience high rotational velocities, these mass estimates remain highly speculative.

\subsection{Spectra of OBe candidates}
 We find that nine of the stars identified as OBe candidates through color cuts are also observed in \citet{lorenzo_new_2022}. For the stars identified through color cuts, most notable are 's003', 's024', and 's132', which show broad emission lines in $H\beta$ and $H\gamma$. We further note that the remaining spectral features in 's003' and 's024' appear relatively shallow, which could indicate high rotational velocity, a typical observable in OBe star spectra. Furthermore, 's045' is noted to have some broad absorption features and emission in H$\beta$, making it a viable candidate in combination with the photometry. 's049' was identified as an emission line star in \citet{Massey07}, but is classified as a supergiant in \citet{Lorenzo22}. This star may be an sgB[e] star; however, only higher resolution spectra may answer this question. 's069' is classified as a potential binary system. We note that H$\beta$ emission is present, and the authors note the possible presence of a hot companion. This is particularly interesting as OBe stars are thought to have stripped star companions, making `s069' a potential candidate for such a system. We find no particularities in the spectra for `s043' and `s073', but that is not uncommon. For some OBe stars, only H$\alpha$ will be in emission, so further spectroscopic or deeper photometric follow-up is needed. Lastly, `s076' and `s079' have low SNR spectra, making further comments difficult. \\
 
%Lastly, we discuss the observed phenomena of the contamination of the BHeb burning sequence in the optical CMD. In \citet{DohmPalmer02b}, they noted that the BHeb to RHeb Sextans A appeared to be a factor of $\sim 2$ larger than expected. Although they note that MS stars may contaminate the BHeb, we find that the estimated $\sim5\%$ underestimates the total contamination. The sole exception is stars with absolute magnitude $M<-5.5$ for which we see an approximate contamination of $\sim 5\%$. However, the contamination reaches $\sim 30\%$ at lower magnitudes. 

To determine the exact fraction, we will need deep H$\alpha$ imaging (e.g., HST-GO-17438, PI: Gull) or spectroscopic follow-up to constrain further binary parameters of these crucial stepping stones in the evolution of massive stars and binaries. 
In addition, a detailed study of contamination levels and their effects on these results should be conducted to explore further star formation histories and stellar population diagnostics, such as the BHeb/RHeb sequence.

\section{OB associations vs. Field Stars} \label{sec:spati}

%In this section, we describe the spatial distribution of our sample of massive stars, identify OB association candidates and runaway candidates, and discuss the stellar age spatial distribution in comparison to literature nebular line studies and neutral H~I studies.

\begin{figure}[ht!]
\centering
\includegraphics[width=\columnwidth]{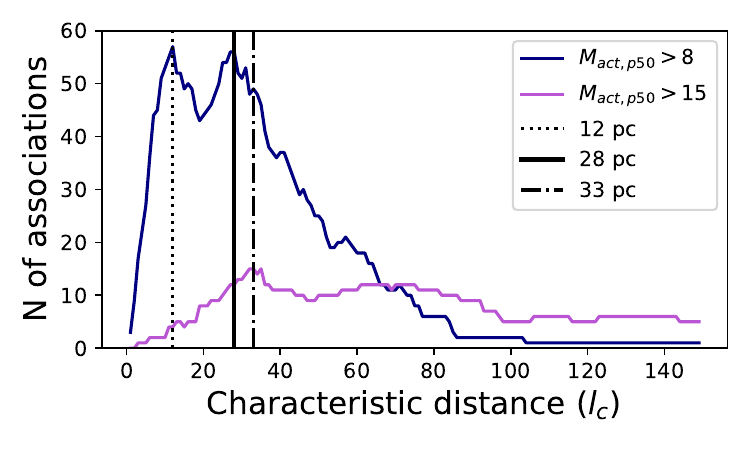} 
 \caption{Number of associations with $N_{\rm stars} \geq 3$ as a function of characteristic distance ($l_{\rm cl}$) for the $M_{\rm act,\,\rm p50} >8$~\msun sample (OB stars; Navy) and for the $M_{\rm act,\,\rm p50} >15$~\msun sample (O stars; Purple). The OB stars appear to show a double peak with maxima at 12~pc and 28~pc. The O stars show their peak at 33~pc.} \label{linklength}
\end{figure}

%\subsection{OB associations vs. Field Stars}\label{sec:field}
We identify OB associations in Sextans A using the friends-of-friends algorithm described in \citet{Battinelli91, Oey04}. We note that, similarly to \citet{Oey04}, we do not distinguish between ``group," ``association," and ``cluster." We briefly summarize the steps here. We use the stars which fall in the following category: $M_{\rm act,\,\rm p50} >8$~\msun, log$(g)_{\rm p50} > 3.5$~dex, and $\chi^2 <100$. This results in 937 stars for which we will try to find associations. In short, the algorithm works by finding all stars ($N_{\rm star}$ within a defined distance (linking length, $l_{\rm cl}$) from each other. 
To determine the best distance, we employ a similar tactic as \citet{Oey04}; we explore a range from 1-150 pc and determine the value $l_{cl}$ for which we recover the most associations. This is our characteristic distance, which is $l_{cl}= 28~{\rm pc}$ for stars with $M_{\rm act,\,\rm p50} >8$~\msun and $l_{\rm cl}= 32~{\rm pc}$ for stars with $M_{\rm act,\,\rm p50} >15$~\msun. We note that for stars  $M_{\rm act,\, \rm p50} >8$ there is a secondary peak at $l_{\rm cl}= 12~{\rm pc}$ (See Fig. \ref{linklength}). 

\begin{figure}[ht!]
\centering
\includegraphics[width=\columnwidth]{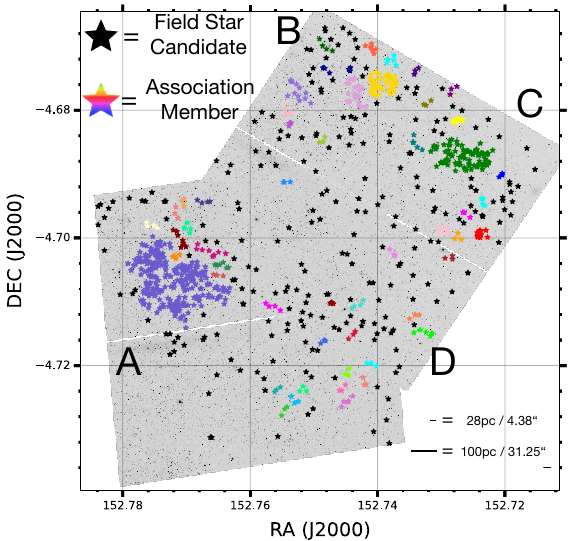} 
 \caption{Mosaic of the Sextans A HST WFC3/F475W footprint overplotted with the stars fitted with the \beast\ that yield; $\chi^2$ $<100$, \logg $>3.5$, and $M_{\rm act}$ $>8$~\msun color-coded by putative association membership. The majority of massive stars are clustered near other massive stars, while $24-28 \%$ appear as field star candidates.
 \label{clustering}}
\end{figure}

This is primarily due to the association in major star-forming complexes (See Fig. \ref{clustering}). In this case, the number of associations increases with shorter lengths because large associations are divided into several smaller ones. The region is likely consistent with several OB associations; however, this requires a more sophisticated study beyond the scope of this paper.
We opt for the characteristic length $l_{\rm cl}= 28~{\rm pc}$ to increase the number of associations outside of H~II regions. We find 57 associations (Table \ref{tab:clust}) , with sizes ranging from $\sim16$~pc to $\sim300$~pc. We note that \citet{Kurtev05} identified 36 OB associations in their search, spanning 30~pc to 120~pc. However, they use ground-based photometry, which may be less deep than our HST photometry and color-magnitude cuts to determine their associations.

Although stars are thought to form in clustered environments \citep{lada03,krumholz19}, it is not uncommon to find massive stars isolated in the field \citep{oey18, DorigoJones20, vargas-salazar20, Gull22, VargasSalazar25}. To determine the fraction of ``isolated" stars, we range the $l_{cl}= 27-29~pc$ to account for stars just at the edge of the currently identified associations. We find that 228-266 stars are "isolated" stars, which is $24-28\%$ of the sample with $M_{\rm act,\,\rm p50} >8$~\msun. We note that a caveat on our method is that we may miss massive stars, which yield high $\chi^2$ values. This effect is at the $2\%$ level in our reported fraction. The derived fraction is consistent with previous predictions by literature \citep[e.g.,]{Oey04} that field stars make up $25-30\%$ of all OB stars in a galaxy. Although a small fraction of these stars could form in situ \citep[i.e., tips of iceberg clusters,][]{Oey04,oey18,vargas-salazar20}, most of these stars are likely runaway candidates. In \citet{lorenzo_new_2022}, sixteen stars were identified as runaway candidates, of which nine overlap with our \hst\ footprint. We find that two stars belong to associations identified in our sample, while seven appear to be isolated candidates. Lastly, we find no significant difference between the fraction of field stars in Sextans A and the SMC. 

\startlongtable
\begin{deluxetable}{cc}
\label{tab:clust}
\tablewidth{0pt}
\tablecaption{OB "Associations" in Sextans A}
\tablehead{ 
\colhead{RA} &\colhead{DEC}}
\startdata
\hline
OBAssoc1& blue \\
\hline
152.720409 & -4.690151 \\
152.720428 & -4.689914 \\
152.720899 & -4.690452 \\
\hline
OBAssoc2& green \\
\hline
152.721927 & -4.686271 \\
152.722678 & -4.686793 \\
152.722680 & -4.686733 \\
152.722852 & -4.686907 \\
152.722869 & -4.686881 \\
152.722877 & -4.686966 \\
152.723091 & -4.687641 \\
152.723297 & -4.687116 \\
152.723446 & -4.686704 \\
... & ... \\
\enddata
\tablecomments{We show here a subset of the catalog; the full catalog is available as mrt.}
\end{deluxetable}

\section{The major star-forming regions in Sextans A} \label{sec:gaslit}

\begin{figure*}[ht!]
\centering
\includegraphics[width=17cm]{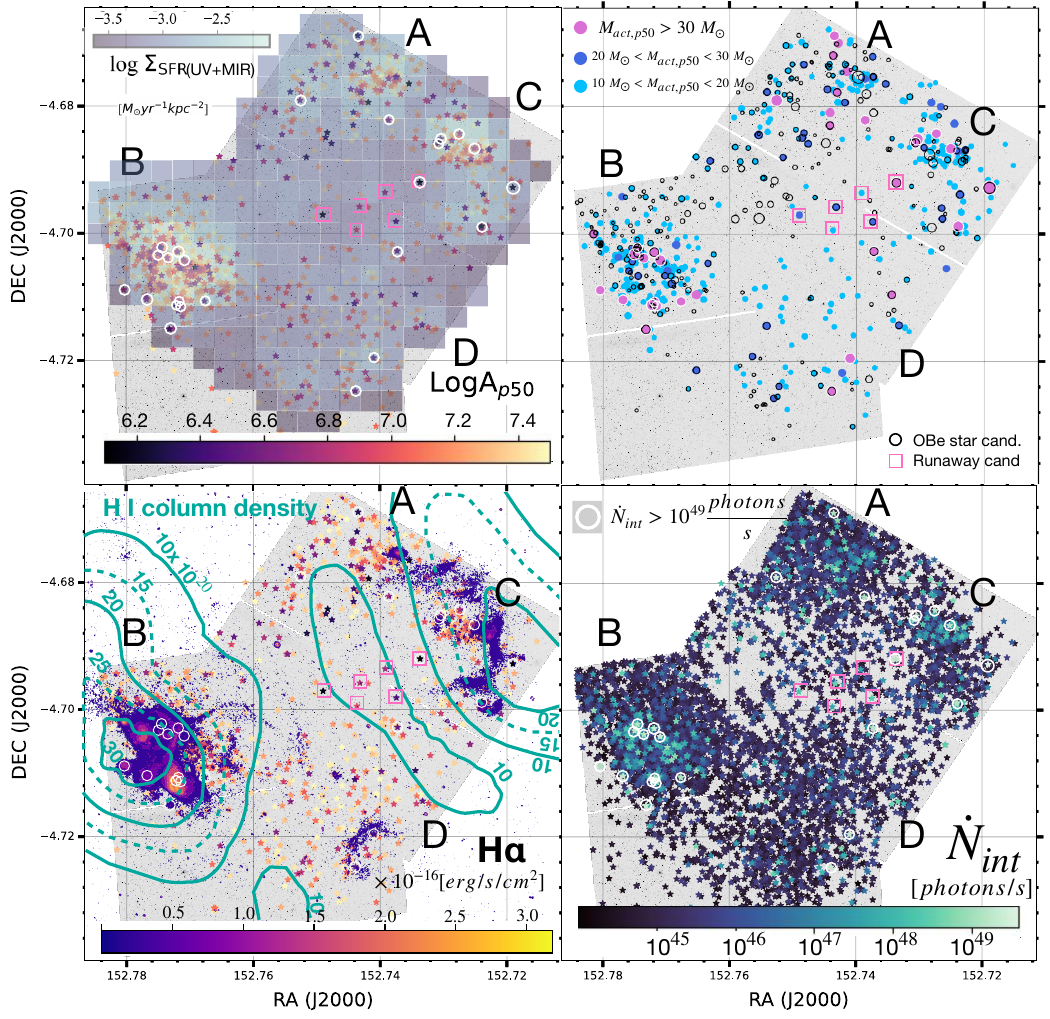} 
 \caption{All panels show a mosaic of the Sextans A HST WFC3/F475W footprint. Top Left: This panel shows the stars fitted with the \beast\ that yield a plausible SED fit ($\chi^2$ $<100$), color-coded by $\log(A)_{\rm p50} < 7.50$. Overplotted is the SFR surface density map in units of \msun yr$^{-1}$kpc$^{-2}$ from \citet{Park25}. The young stellar population coincides with the highest SFR density areas, as is expected. However, near region A and towards the center of the galaxy, young stars appear in low-SFR-density regions. A subset of the stars furthest from the star-forming regions is discussed as runaway candidates (pink squares). The stars with the highest predicted intrinsic ionizing rate ($N_{int}>10^{49}[photons/s]$, white circles) do largely overlap with the highest density SFR regions (a few outliers), and are among the youngest stars in the sample $\log(A)_{\rm p50} < 6.7$~dex.
 Top Right: This panel shows the stars fitted with the \beast\ that yield a plausible SED fit ($\chi^2$ $<100$), color-coded by their current mass ($M_{\rm act,\,\rm p50}$). OBe star candidates with approximate mass ($M_{\rm act,\,\rm p50}>8$\msun, WFC3/F275W $<22.9$) are indicated with a black circle. Some of the massive OBe candidates yield $\chi^2$ $<100$, despite showing IR-excess. Therefore, the masses of these stars need to be interpreted with caution. The distribution of mass traces the star-forming regions, where the most massive stars are clustered near the H~II regions. Three of the six runaway candidates are also OBe star candidates, including a $M >30$ \msun candidate. 21 of the 34 stars with $\chi^2$ $<100$ and $M >30$ \msun appear as bona fide $M >30$ \msun candidates. 23 of them provide a predicted intrinsic ionization rates of $N_{int}>10^{49} [photons/s]$. 
 Bottom Left: Similar to the top panel, this shows the stars fitted with the \beast\ that yield a plausible SED fit ($\chi^2$ $<100$), color-coded by $\log(A)_{\rm p50} < 7.50$. Overplotted are contours of the H~I column density (green) as observed by \citet{Skillman88} as well as the intensity map of the H$\alpha$ emission (purple-yellow) from ground-based imaging (Local Volume Legacy Survey; \citealt{kennicutt07}). Region A is void of H$\alpha$ emission, despite hosting young and massive stars, including two stars with predicted intrinsic ionization rate of $N_{int}>10^{49} [\rm photons/\rm s]$. 
 Bottom Right: This panel shows the stars fitted with the \beast\ that yield a plausible SED fit ($\chi^2<100$, no age restriction), color-coded by their \beast\ predicted intrinsic LyC photon ionization rate (only stars with $\dot{N}_{int}>10^{44}$ [photons/s]). The sources with the highest predicted intrinsic LyC ionizing photon rate are indicated with white circles. In theory, these should power a strong H$\alpha$ emission. This is true for the majority of the sources, but five sources seem not to power any H~II emission, including one of the runaway candidates.}
\label{sfrHaNintmap}
\end{figure*}

 \begin{figure*}[ht!]
\centering
\includegraphics[width=\textwidth]{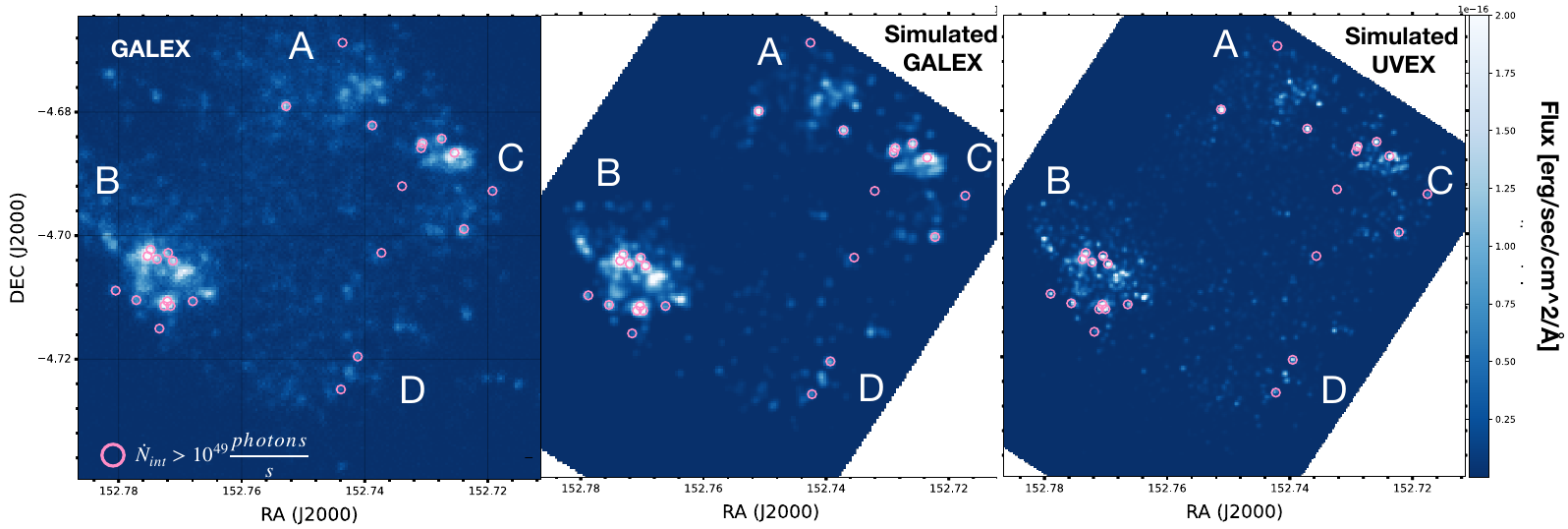} 
 \caption{Left: FUV image of Sextans A observed by GALEX with the AIS filter. Similarly to Fig. \ref{sfrHaNintmap}, the highest flux density regions overlap with the young stellar population of Sextans A and the regions with the highest ionization. Middle: Simulated GALEX AIS image using the \beast\ results for stars that yield a plausible SED fit ($\chi^2$ $<100$). The simulated image reproduces the observed data, suggesting that the dominant sources of the FUV image are well reproduced by the current \beast\ fits to the hot stars at GALEX resolution and sensitivity. Since we apply a magnitude cutoff of F475W$<24$, the simulated image does not reproduce the UV noisy background seen in the GALEX image. The stars with the highest predicted intrinsic ionizing rate ($N_{int}>10^{49}[\rm photons/\rm s]$, pink circles) largely overlap with the highest FUV flux region. In some instances, the predicted intrinsic ionizing rate is high, whereas the observed FUV flux is not, most likely due to extinction. 
 Right: Simulated UVEX FUV image using the \beast\ results for stars that yield a plausible SED fit ($\chi^2$ $<100$). Even with the increased resolution, the densest star clusters in the H~II regions remain unresolved. Nevertheless, compared to GALEX, UVEX will resolve stars to individual sources in most regions of the Sextans A and will have increased sensitivity to fainter FUV sources (e.g., stripped stars). Additionally, the prospect of obtaining multi-epoch photometry of Sextans A would provide crucial insights into the variability and multiplicity of massive stars in low-metallicity environments.
 \label{uvmap}}
\end{figure*}

Sextans A is known to host four major star-forming complexes and several H~II regions  \citep[see Fig. \ref{sfrHaNintmap}, e.g.,][]{Hodge94,Magrini05,garcia19a}, and is known to be a gas rich system in terms of high H~I surface density as well as large ratio of neutral hydrogen mass to optical luminosity \citep{Skillman88}. The star-formation mainly occurs on the outskirts of the galaxy, while the center is characterized by a large H~I deficiency (see Fig. \ref{sfrHaNintmap}). \citet{Warren_HI_2011} concluded that the feedback from recent central star formation was sufficient for creating the central H~I hole in Sextans~A.

In this section, we discuss the results of our SED fitting (in particular, stellar ages and predicted ionizing flux) in the context of literature nebular line analysis of the H~II regions and the neutral H~I gas in Sextans A.
Figure \ref{sfrHaNintmap} shows four composite HST images of the LUVIT Sextans A observations, overplotted with \beast\ results: a star-formation rate (SFR) map derived from UV and MIR observations \citep[100 Myr timescale,][]{Park25}; an H$\alpha$ map \citep{kennicutt07}; H~I column density contours \citep{Skillman88}; and GALEX FUV imaging. Young stars ($\log$A$ < $7.5) are color-coded by their age from our SED fitting. For clarity of discussion, we split the discussion into four major star-forming regions \citep{vanDyk98, DohmPalmer02,garcia19a} within Sextans A \citep[A, B, C, D, see Fig. \ref{sfrHaNintmap}, and][]{DohmPalmer02,garcia19a}. Figure \ref{uvmap} shows the GALEX FUV imaging of Sextans A, as well as simulated GALEX FUV and UVEX imaging created using the  \beast\ results.

\subsection{Region A:}
\citet{DohmPalmer02} first identified this association of star-forming regions and describes it as the oldest of the star-forming regions with star-formation ongoing for at least 400 Myr or longer. It was also pointed out that the gas in this region is depleted and would not likely be able to form new stars for much longer (See Fig. \ref{sfrHaNintmap}). H$\alpha$ emission observations of this region were taken as part of the Spitzer Local Volume Legacy Survey \citep[LVL][]{kennicutt07,LVLIRSA}, and the ionized gas was studied in \citet{Gerasimov22}. In this paper, no H$\alpha$ emission was observed, in line with the H$\alpha$ emission map from LVL (Fig. \ref{sfrHaNintmap}). Lastly, ten stars (SpT B2-O8) in the complex were analyzed as part of \citet{lorenzo_new_2022}. 

Our \beast\ fits reveal that this region harbors a decent young stellar population (200 stars with $\log A < 7.5$). We find 96 massive MS stars ($\chi^2<10$) in this region and identify three sources with high ($\dot{N}_{int}>10^{49} [\rm photons/\rm s]$) LyC photon ionization rates. The presence of young hot stars is further supported by the UV flux observed in GALEX, which is well reproduced in our GALEX simulated image using the \beast\ results. A high H$\alpha$ flux over-density is observed right at the transition of the H~I $\sim 10\times 10^{20} \rm cm^{-2} $ column density contour. This does coincide with a higher stellar mass density compared to the remaining stellar mass distribution in this star-forming complex. 

As suggested by previous studies, star formation is still ongoing in this region despite the overall lack of observed ionized gas. However, the \beast\ ages suggest star-formation as recent as $\sim 2~\rm Myr$ ago. The masses suggest stars with $M_*>10$ \msun, implying these stars are unlikely to have formed more than $\sim$40 \rm Myr ago. Additionally, several high LyC photon-ionization-rate producers are present in this area. This seems contradictory to the lack of H$\alpha$ emission in this region. Although the H~I column density in this region is lower than in other regions of the galaxy (``only" $5\times10^{20} \rm cm^{-2}$), this should be sufficient to have produced ionized gas given the stellar population presented here. But given the diffuse distribution of the stellar population, it is plausible that the lack of observed H$\alpha$ emission in this region is due to the limited depth of ground-based H$\alpha$ observations.

\subsection{Region B:}
Region B is identified by \citet{DohmPalmer02} as the second-oldest region, which would have been star-forming for at least 200 Myr. In contrast to Region A, the region is embedded in a high H~I column density blob, with the peak of star-formation being slightly offset from the densest gas region (See Fig. \ref{sfrHaNintmap}). LVL photometric observations of the H$\alpha$ emission exist as well as narrow-band photometric and spectroscopic observations of the ionized gas \citep{Gerasimov22}. The latter observations find that the region consists of several dominant H~II regions, surrounded by faint, extended filaments \citep{Gerasimov22}. They also estimated $Q^0_{stars}= 2.42 \times 10^{50}$ and derived $Q^0_{H_{\alpha}}= 1.08 \times 10^{50}$ for complex B, suggesting an escape fraction of 0.55. Lastly, the authors identified signs of ionized gas outflow, as evidenced by significant differences between the kinematics of ionized and atomic gas. \citet{lorenzo_new_2022} analyzed 61 stars (SpT O3-B9) in the complex, including two of the earliest spectral type stars (s001 and s002). Lastly, polycyclic aromatic hydrocarbons (PAHs) were detected in Region B \citep{Tarantino25}. 

Unsurprisingly, the \beast\ fits reveal a high stellar density of young stars coinciding with the highest density regions of the star-formation density map. The region harbors 704 young stars and 196 massive stars, of which 14 have masses above 30 \msun. The region is likely composed of several nearby associations, groups, or clusters, although they are identified as a single large association in our previous section (See Fig. \ref{clustering}). The region harbors eleven sources with high ($\dot{N}_{int}>10^{49} [\rm photons / \rm s]$) LyC photon ionization rates, with three of them appearing particularly close together in the southern region of the complex. These three sources power the highest intensity H$\alpha$ emission in the region and are three of the youngest stars ($\log(A)\sim 6.3-6.6$). In the GALEX image, the region appears as the brightest and most extended FUV region. The simulated image reproduces the general structure of the UV emission; however, the two sources in the north-east region of the complex appear brighter in the simulated image. These stars are OBe candidates, implying that the \beast\ is likely not capturing the UV attenuation of the disk in the current fit, resulting in the brighter UV signal in the simulated image.  

The ionized gas appears well traced by the stellar density and mass distribution, except for one H$\alpha$ emission region in the north-east part of the galaxy. The \beast\ does suggest at least one young O-type star in the region, which is identified by \citet{lorenzo_new_2022} as s026, which the authors report to be an O8V type star. It is, however, surprising that the star would provide this intense H$\alpha$ emission, especially in comparison to the hotter, more massive stars in the same region. For example, the western region of the complex is populated by a similarly dense massive star population; however, we do not observe the same H~I column density as in the north-east part of the galaxy.  

We briefly comment on the location of the PAH clumps in comparison to the sources with the highest LyC photon ionization rates. In particular, in the south-west region of the complex, two clumps (12, 13) are found that are in close proximity ($<5~ \rm pc$ for 12, $<15~ \rm pc$ for 13) to three of the highest ionizing sources and to the most massive stars in the galaxy. The high column density in this region likely provides sufficient UV shielding to prevent PAH erosion \citep{Thatte26}.  

Lastly, it is evident that GALEX cannot resolve the individual UV sources in this region due to the instrument's insufficient resolution relative to the crowding in the area (See Fig. \ref{uvmap}). The improvement in UVEX compared to GALEX is perhaps most evident here. Although the region remains semi-resolved, the individual UV sources are more easily distinguished from each other, providing much-needed spatial information that will aid in constraining the FUV brightness of various sources in this region. Ultimately, fully resolving this crowded region will require a telescope with FUV capabilities and a resolving power comparable to or better than HST.

\subsection{Region C:}
Region C is identified by \citet{DohmPalmer02} as the youngest region in Sextans A ($<20$ Myr). Similar to Region B, this star-forming region is in a high H~I column density area, and there is also an offset from the densest gas region and the star-formation peak (See Fig. \ref{sfrHaNintmap}). LVL photometric observations show dispersed H$\alpha$ emission with a high intensity region at the center of the region.
In \citet{Gerasimov22}, the region is described as having a strong offset between the kinematics of the atomic and the ionized gas. In particular, H~I appears to be more perturbed, perhaps explaining the discrepancy, which suggests that the emission of the atomic and ionized gas may come from different regions along the line of sight \citep{Gerasimov22}. They derived $Q^0_{H_{\alpha}}= 7.57 \times 10^{49}$ and estimated $Q^0_{stars}= 1.78 \times 10^{50}$ for complex C, suggesting an escape fraction of 0.58. Lastly, \citet{lorenzo_new_2022} analyzed 14 stars (SpT O5-B9) in the complex.

Compared to Region B, the \beast\ fits reveal a much smaller area of high stellar density in the northern part of the region, while some young massive stars are found in the southern part of the region (267 stars with $\log A < 7.5$, and 146 massive stars). The star-formation density peaks in the northern part, coinciding with most of the massive stars and the majority of the sources with high ($\dot{N}_{int}>10^{49} [\rm photons/ \rm s]$) LyC photon ionization rates in this region. We find three stars with suggested \beast\ masses larger than 30 \msun. However, the star in the westernmost region, appearing as the most massive star, is also tagged as an OBe star candidate, hence the mass may be lower (See Sec. \ref{sec:bestars}). In the GALEX image, the region is very well recovered by the \beast\ simulation, even the OBe star candidate in the westernmost region.

The H~I column density in the region is the second highest in the galaxy, and the peak density is offset to the west of the peak star-formation density. Similarly, the bulk of massive stars in the region do not overlap significantly with the region of highest atomic gas density. When looking at the ionized gas, it appears that there is also an offset from the bulk of massive stars compared to the highest peak in H$\alpha$ emission in the region. In particular, the highest sources of LyC photon ionization only partially overlap with the highest H$\alpha$ emitting regions. \citet{Gerasimov22} previously noted the complex gas kinematics in this region. To better understand these kinematics and their interplay with the massive star population, further spectroscopic follow-up is necessary, preferably with an IFU. 

\subsection{Region D:}
Region D was first identified by \citet{garcia19a} through the discovery of three O-type stars and the highlighting of a faint rim of ionized hydrogen. Unsurprisingly, this region was detected later than the others, since the gas is more diffuse and the massive stars are sparse and dispersed. \citet{garcia19a} observed four O-type stars in the region, which were also re-observed by \citet{lorenzo_new_2022}, in addition to seven OB-type stars. We note that we discuss s3/s029 in detail in Sec. \ref{gracestar}. \citet{Gerasimov22} analyzed the region as part of their study and reported a high velocity dispersion, particularly around s4/s004. Thus far, it has been hypothesized that this could be caused by a local wind-driven bubble or by stellar ejecta. The former is difficult to explain given current observational evidence on wind strength in this metallicity regime \citep{telford_observations_2024,Furey25}. Lastly, they derived  $Q^0_{H_{\alpha}}= 1.18 \times 10^{49}$ and estimated $Q^0_{stars}= 3.75 \times 10^{49}$ from \citet{Gerasimov22} and an escape fraction of 0.69.

This region appears the most sparse and dispersed in terms of massive stars for our \beast\ fits as well (113 stars with $\log A < 7.5$, and 36 massive stars). We do find three massive stars with masses greater than 20 \msun, identical to stars s2, s3, and s4 in \citet{garcia19a}. While the authors suggest that one star may have formed in isolation, we find that each star has at least two other nearby stars with masses greater than 8 \msun. While this may still have implications for the IMF, either because it is stochastically sampled or truncated, these stars do appear to have formed in an association. Investigating the IMF requires a more complete study of the underlying stellar population in this region, which is beyond the scope of this paper. 

The two most massive and youngest stars (s2,s4) also have the highest LyC photon ionization rate, which likely powers the dim H$\alpha$ emission, despite being in a low atomic gas density region. We note that s4 appears to be one of the most massive stars in the entire galaxy. The mass of s2 remains more uncertain as it is one of our brightest OBe star candidates. We note that s3, the third most massive star in the region, is known to have a weak stellar wind \citep{telford_observations_2024,Furey25}. Based on the \beast\ mass, this is perhaps unsurprising as the star is in the weak-wind regime. While we cannot further comment on the strength of the winds of s2 and s4 without spectra, it is interesting to consider the stellar ejecta as a possible source for the dispersion observed by \citet{Gerasimov22}. In particular, s4 is marked as a binary candidate by \citet{lorenzo_new_2022}, while s2 is identified as an OBe star candidate in this study. If indeed these stars are binaries or binary products, the ejecta may result from non-efficient mass-transfer \citep{Izzard13,CournoyerCloutier25}. Multi-epoch spectra, UV spectra, and spatially resolved ISM maps are needed to constrain the various hypotheses present in the literature and this paper. 

Lastly, this region is perhaps one of the most promising regions to target with UVEX. As shown in the simulated image (Fig. \ref{uvmap}), all three massive stars are well resolved and represent the brightest FUV sources. This would provide a unique opportunity to study some of the most massive stars in one of the lowest-metallicity environments in greater detail. Obtaining UV spectroscopy would reveal important stellar-wind features or even potentially hot companions, such as stripped stars. In synergy with the nebular line observations, this would ultimately provide crucial constraints for the interpretation of the UV regime routinely observed at high redshift.  

\subsection{Field}
The remaining massive stars appear to be field star candidates, though a fraction may originally have formed in associations and have since dispersed. A subset of the field stars are likely runaway stars, i.e., those ejected from their birth cluster due to dynamical interactions \citep{oey18,vargas-salazar20}. These mechanisms primarily rely on binaries driving ejections through supernova explosions of one of the companions (binary supernova scenario) or close encounters with a binary system (dynamical ejection scenario). 
 
We discuss only six candidates here, the most likely runaway candidates based on their \beast-derived ages and positions in Sextans A. We note that this sample overlaps with only one of the isolated OB stars in \citet{lorenzo_new_2022} sample. The sample in \citet{lorenzo_new_2022} includes eighteen stars, nine of which are within our footprint. When comparing the sample with the results from the \beast, it appears that seven are massive in our \beast\ results and eight have nearby OB-type stars within a characteristic distance, which is why we do not include these in the discussion of runaway candidates.

The six stars in our sample all have derived ages $\log(A)< 7$ and masses ranging from 11-23 \msun, with the youngest star ($\log(A)=6.14$) at the center of the galaxy with a mass of $\sim 21$ \msun.  Attempting to associate each with their closest star-forming region or "association," we find distances ranging from $150-350$~pc. This would suggest that the stars need velocities ranging from $\sim 50$ \kms $ - 340$ \kms ($50$ pc/Myr $- 350$ pc/Myr) to reach their current position in Sextans A within their lifetime. These velocities are in line with predictions for runaway stars, and appear to be generally too high to be due to the natural dispersion of the galaxy. 
Except for one of the candidates, the current ground-based H$\alpha$ and GALEX imaging do not indicate the presence of hot stars, despite all stars having temperatures $>30 ~\rm kK$. Perhaps this lack of UV luminosity can be explained by the extinction due to relatively high $A_V$ reported by the \beast\ in this region or due to their circumstellar disks in case of the OBe candidates. We find a median value of $A_V \sim 0.5$ and a range of $A_V \sim 0.2-0.7$ reported by all stars fit in this region. The median is more than twice as high as the median throughout the entire galaxy, which is around $\sim 0.22$~dex, with an 84th percentile of $\sim 0.44$~dex. 
This reported extinction could explain why no bright UV luminosity is observed with GALEX. To show this, we simulated the anticipated FUV intensity using our \beast\-fits including and excluding $A_V$ (See Fig. \ref{uvmap}). The lack of H$\alpha$ emission is unsurprising as the H~I column density in the region is low ($<10^{21} \rm cm^{-2}$), and a single star will not ionize the region enough to produce significant H$\alpha$ emission detectable at the LVL limit. 

Of particular interest is the candidate in the upper right near region C. It is a massive OBe star candidate, although the actual mass is likely uncertain due to its classification as such (See Sec.\ \ref{sec:bestars}). Nevertheless, the star appears as an isolated UV source in the GALEX image, albeit quite faint. Additionally, the SED fit suggests a high ($\dot{N}_{int}>10^{49} [\rm photons/ \rm s]$) intrinsic LyC photon ionization rate. 
 
Lastly, we find some sources of high H$\alpha$ intensity not near any H II region. While this is speculative, common sources of strong H$\alpha$ emission not associated with extended gas regions are binary interactions and binary products (e.g., OBe stars and stripped stars). Explaining the sources of ionizing radiation and UV luminosity at high redshift remains challenging, partly because of limited local empirical constraints. Sextans~A, and a handful of other sub-10\% \zsun LG galaxies, are some of our best local test-beds for these faraway galaxies.
 
As evident in Figure \ref{sfrHaNintmap}, the distances to these galaxies prevent shallow ground-based observations (H$\alpha$) or low-resolution instruments (GALEX FUV) from capturing more than the brightest features. From a photometric perspective, HST photometry is crucial for characterizing these stars, especially in the UV (See Appendix \ref{sec:UV}). Future UV-sensitive space-based missions, such as UVEX or the Habitable World Observatory (HWO), if the UV capabilities are supported, would provide much-needed tools to tackle these outstanding questions at high redshift.

\section{The LyC escape fraction of Sextans A} \label{escapef}
Massive stars in galaxies throughout the epoch of reionization are leading candidates for cosmic reionization through their LyC radiation. Recent JWST results have strengthened this hypothesis by finding high-redshift galaxies with unexpectedly bright UV luminosities and surprisingly strong Ly$\alpha$ emission \citep[e.g.,][]{Chisholm20,Chisholm22,Llerena24,Giovinazzo25}. These measurements are mostly through indirect tracers, where high $f_{\rm esc}$ has been linked to high SFR, low metallicity, strong nebular emission lines, highly ionized gas, low dust, strong Ly$\alpha$ emission \citep[e.g.,][]{Bouwens10,Pellegrini12,Zackrisson13,Alexandroff15,Steidel18,Choustikov24,Carr25,Jaskot25}. While local galaxies are not perfect analogs of high-r,edshift ones, they remain the only laboratory where the physical mechanisms governing LyC photon escape can be spatially resolved and dissected in detail. Consequently, local low-metallicity star-forming galaxies provide the critical constraints needed to interpret unresolved high-redshift observations. Motivated by this, we utilize the \beast\ and the LVL map to quantify the escape fraction as outlined by \citet{choi20}. We perform this analysis first on individual star-forming regions to understand the local physics, and subsequently integrate over the entire galaxy to simulate the unresolved perspective. Further details can be found in \citet{choi20}.
Essentially, the escape fraction can be described as follows: 
\begin{equation}
f_{\rm esc} = \frac{\dot{N}_{\rm int}-\dot{N}_{\rm gas} -\dot{N}_{\rm dust}}{\dot{N}_{\rm int}}
\end{equation}
, where $\dot{N}_{\rm int}$ is the intrinsic rate of production of
ionizing photons, $\dot{N}_{\rm gas}$ is the rate of consuming ionizing photons by neutral hydrogen from the H$\alpha$ recombination line, and $\dot{N}_{\rm dust}$ is the rate of absorption of ionizing photons by dust. 

First, we use the \beast\ results to sum the intrinsic ionizing photon production rate of all stars in the defined region to obtain $\dot{N}_{\rm int}$. Then we estimate $\dot{N}_{\rm dust}$, which is difficult to directly measure as we are limited to line-of-sight extinction \citep{choi20}. We use the following formula:
\begin{equation}\label{eq2}
\dot{N}_{\rm dust} = f_{\rm cov}\times(\dot{N}_{\rm int}-\dot{N}_{\rm gas}) 
\end{equation}
, where $\dot{N}_{\rm gas} = 7.315\times10^{11}\times L(H_{\alpha})$ and $f_{\rm cov}$ refers to the dust “covering factor” \citep[i.e., the fraction of the H~II region surface where dust absorbs LyC photons,][]{choi20}. To calculate $L(H_{\alpha})$, we sum the calibrated H$\alpha$ flux from the LVL image, correct the flux for dust, and then convert it into luminosity using the distance of Sextans A. To correct for the dust, we calculate the Balmer decrement assuming Case B recombination. We use the values for the H$\alpha$ flux and H$\beta$ flux in \citet{Magrini05}, which exist for three points in regions B and C. The mean value of $E(B-V)$ for region B is $0.04$, and $0.02$ for region C. We adopt a mean value of $E(B-V)=0.03$ for regions A and D, since no measurements of both the H$\alpha$ and the H$\beta$ flux are reported in the literature for these regions.

Using Eq.\ref{eq2}, $f_{\rm cov}$ is also used to simplify the complex ISM geometry, since we are only able to determine the attenuation in our line-of-sight.  As in \citet{choi20}, the gas/dust geometry assumes a dust-free H~II cavity enveloped by a porous photo-dissociation region. In general, $f_{\rm cov}$ is the most uncertain factor in this calculation without independent, comprehensive modeling of star formation, including dust emission \citep[e.g.,][]{Hermelo2013}. 

We therefore explore a range of porous geometries by adopting $f_{\rm cov}$ = {0.2, 0.4, 0.6} to calculate the LyC escape values. These intermediate to low $f_{\rm cov}$ values are motivated by the steep decline in the dust-to-gas ratio at low metallicities like Sextans A \citep{Inoue01}, the prevalence of density-bounded conditions driven by hard ionizing spectra \citep{Jecmen23}, and the reddening to neutral gas covering fraction relation derived in \citet{Reddy16}. 

Lastly, to determine the escape fraction of the entire galaxy, we sum the intrinsic ionizing photon production rate of all stars in our catalog and similarly use the same footprint to extract $L(H_{\alpha})$. We summarize the derived values in Table \ref{tab:esfr}. 

As mentioned in the previous section, \citet{Gerasimov22} derived values for the escape fraction in their study of Sextans A by using \citet{lorenzo_new_2022}'s spectral typing of OB stars in each star-forming region to match them with models from \citet{martins05} to determine the total amount of the hydrogen ionizing photons. They then use their Kitt Peak National Observatory H$\alpha$ imaging to determine the total amount of the hydrogen ionizing photons required to produce the observed H$\alpha$ flux.
Compared to the values derived in \citet{Gerasimov22}, we find that the values for region B ($0.55$) and region C ($0.58$) are within our range of suggested values. However, for Region D, their value ($0.69$) would require us to assume a near-zero $f_{\rm cov}$ ($<0.01$). Even with Sextans A's general low dust, this assumption would be hard to physically motivate.

\startlongtable
\begin{deluxetable}{cccccc}
\label{tab:esfr}
\tablewidth{0pt}
\tablecaption{Predicted LyC photon ionization rates and the escape fraction }
\tablehead{ 
\colhead{Region} &\colhead{$\dot{N}_{int}$} & \colhead{$\dot{N}_{gas}$}&\colhead{$f_{\rm esc}$}&\colhead{$f_{\rm esc}$}&\colhead{$f_{\rm esc}$}\\
\colhead{} &\colhead{phot./s} & \colhead{phot./s} &\colhead{$f_{\rm cov}=0.2$}&\colhead{$=0.4$}&\colhead{$=0.6$}}
\startdata
A&$3.5\times10^{50}$&$1.7\times10^{49}$&$0.76$&$0.57$&$0.38$\\
B&$7.4\times10^{50}$&$1.8\times10^{50}$&$0.61$&$0.46$&$0.31$\\
C&$6.2\times10^{50}$&$8.0\times10^{49}$&$0.70$&$0.52$&$0.35$\\
D&$6.0\times10^{49}$&$2.12\times10^{49}$&$0.52$&$0.39$&$0.26$\\
Global&$2.6\times10^{51}$&$3.1\times10^{50}$&$0.71$&$0.52$&$0.35$\\
\enddata   
\end{deluxetable}

The LyC escape fraction in Sextans A is consistently high and displays variations across the galaxy ($0.26-0.76$) with our assumed $f_{\rm cov}$ values. The global escape fraction ranges from $0.35-0.71$. This high escape fraction exceeds the values typically required for cosmic reionization, which are of order $f_{\rm esc}\sim0.1$–0.2 \citep[e.g.,][]{Robertson15}. To achieve an escape fraction below $f_{\rm esc}<0.2$, we would need to increase $f_{\rm cov}$ to $\sim0.8$. Given the low dust content in Sextans A, this would be surprising. More recent simulations and studies suggest that $f_{\rm esc}$ should evolve with UV luminosity or stellar/halo mass \citep[][]{Ma20,Robertson21,Rosdahl22,Jaskot25}.
That being said, the global $f_{\rm esc}$ of Sextans A remains high. It is notable that Sextans A has a stellar mass of $\sim 4.4 \times 10^7$ \citep{Skillman88}, which is close to the stellar mass at which $f_{\rm esc}$ peaks across high-redshift ($z>6$) galaxies in the \texttt{SPHINX} simulations \citep[$M_*\sim10^7$][]{Rosdahl22} and the \texttt{FIRE-2} simulations \citep[$M_*\sim10^8$][]{Ma20}, albeit at lower overall $f_{\rm esc}$ values. Additionally, the low column density across Sextans~A also supports the \texttt{FIRE-2} simulations' suggestion that high-$f_{\rm esc}$ stars are located in regions with lower neutral and ionized gas lines. 
Overall, Sextans A may not be identical to the earliest galaxies, but could be consistent with an intermediate-mass, low-metallicity star-forming galaxy, as suggested in \citet{Ma20}, that starts to dominate toward the end of reionization at $z \sim 6$.

While these results provide a first benchmark for galaxies with metallicities lower than that of the SMC, certain shortcomings should be noted. First, the impact of dust parameters such as the opacity, size, and amount of the dust could significantly impact the escape fraction. Additionally, the dust and gas configurations in Sextans A may vary from star-forming region to star-forming region \citet{Calzetti01} and be subject to complex geometries in Sextans A, as mentioned in \citet{Tarantino25}. Second, we currently do not consider binaries or rotation, which may affect the SEDs used to calculate $\dot{N}_{\rm int}$. Lastly, current studies of the ionized gas remain relatively shallow. A detailed map of the ionized gas in Sextans A would provide significant empirical constraints on the interaction between massive stars and the ISM. With the exception of dust, most of these would actually lead to an increased $f_{\rm esc}$ \citep{choi20,Ma20} value. Further dedicated studies of dust, stars, and gas are necessary to fully capture the intricate relationships among these factors, especially to inform the theoretical models widely used to explain high-redshift observations of low-metallicity environments. 

\section{Conclusion}
We present a panchromatic photometric characterization of the metal-poor massive star content in Sextans A. Our findings from the LUVIT observations are as follows:

\begin{itemize}

 \item We analyze the bright stars ($F475W < 24$ mag, 14874 stars) in Sextans A using the LUVIT \hst\ photometry. We perform SED fitting using the \beast\ for all stars in the sample.

 \item Using the spectral-typing in \citet{lorenzo_new_2022}, we compare the \beast\  derived temperatures versus spectral type to the theoretical prediction of the relation as derived in \citet{Lorenzo25} and the observed relation in the SMC of \citet{Dufton19}. We find that for the 22 dwarf-type stars, our derived parameters agree with the relationship, except for four outliers that appear to show signs of binarity, which may impact the \beast\ fit. However, they seem to trend towards slightly lower temperatures overall than predicted by the relationship. For the 32 (sub) giant-type stars, we find eight stars. Two stars have low-resolution spectra, which may make spectral typing less reliable; one star appears to be in a binary, and one shows no apparent signs to explain the discrepancy. The remaining four stars seem to be classified as supergiants in \citet{lorenzo_new_2022}. Supergiants are currently not well captured by the \beast, which is likely the source of disagreement in this case.
 
\item We compare the identified massive stars to existing literature on the massive star content in Sextans A \citep{camacho16,garcia19a,telford_far-ultraviolet_2021,telford_observations_2024}. Most spectroscopically observed stars that also yield a reliable $\chi^2$ value for their photometric fit show good agreement (i.e., fall on expected locations in the HR diagram and CMD for given parameters and spectral type). In most cases, disagreements can be explained by binarity, low spatial resolution, or low signal-to-noise spectra. We investigate the star s3/s029 in greater detail and rule out the possibility that the flux contribution from unresolved nearby sources in the spectroscopy drives the disagreement between the spectroscopic and photometric masses. 
  
\item We show indications that the OBe star fraction increases at low metallicity. We obtain OBe star fractions of $15\%$, $23\%$ and $17\%$ above 8 \msun, 15 \msun\ and 20 \msun, respectively. We comment on the eleven brightest and likely most massive OBe star candidates. A comparison with spectra suggests that most of these stars are indeed OBe stars.

\item We identify 57 OB associations in Sextans A using a friends-of-friends algorithm with a characteristic linking length of 28 pc, and find that $24–28\%$ of massive stars ($M>8, M_\odot$) are isolated, consistent with measurements of the isolated massive stars in the SMC.

 \item We compare the results from the \beast\ to prior literature on the four major star-forming complexes in Sextans A. Overall, the young ($\log (A)<7.5$) stellar population aligns with the GALEX FUV map and H$\alpha$ emission, even where H$\alpha$ is weak or offset.
 We identify 96 massive stars in region A, 196 in region B, 146 in region C, and 36 in region D, including multiple sources with high intrinsic LyC photon rates ($\dot{N}_{\mathrm{int}} > 10^{49}$). Outside the main complexes, we find a small population of young massive stars ($\log A < 7, 11–23 M_{\odot}$) that appear dispersed across the galaxy. We identify six runaway candidates with velocities ranging from $\sim 50$ \kms $ - 340$ \kms based on their age and location in the galaxy. These stars show higher extinction and weak or absent H$\alpha$ and FUV signatures in GALEX imaging, consistent with their isolation and local conditions.

\item We find that Sextans A exhibits consistently high LyC escape fractions, with global values of $f_{\rm esc}=0.35,0.52,0.71$~ (for $f_{\rm cov}=0.2,0.4,0.6$) and regional values ranging from $0.27$ to $0.76$. This supports the idea that intermediate‑mass, metal‑poor dwarf galaxies may contribute significantly to reionization during brief phases. This suggests that intermediate‑mass, low‑metallicity systems like Sextans A can leak ionizing photons at levels exceeding those required for cosmic reionization. We highlight the need for improved constraints on dust geometry, stellar evolution pathways, and the distribution of ionized gas.
 
 \end{itemize}

%This study highlights the utility of panchromatic photometry in characterizing the general population of metal-poor massive stars in nearby dwarf galaxies. While photometry cannot replace medium - to high-resolution spectra, it is a cost-effective alternative for identifying the highest-priority targets for different science goals. Future work will include further follow-up on the diverse outliers mentioned in this work (e.g., OBe stars, binaries, and stripped stars), as well as explorations of the evolved stellar population. 

\appendix

\section{Uncertainties in the \beast\ fitting}
We simulate two sets of stars with $M>5$~\msun based on our Sextans~A observations and noise model and the ASTs to (1) quantify the effects on the fitting in the absence of UV photometry and (2) the effects of fixing metallicity, distance, and extinction curve (Rv). The first catalog is generated with fixed distance, metallicity, and extinction curve (Rv). The second catalog varies distance, metallicity, and $R_V$. Using the \beast, we can generate these simulated catalogs and fit them using the same infrastructure as our regular fits. In both cases, we fit the simulated data using the same setup as for the analysis (Sec. \ref{sec:beast}) presented in the paper.

\subsection{The Importance of the UV}\label{sec:UV}
The UV is particularly sensitive to the properties of massive MS stars. They have typical temperatures of ($>15 $~kK) and therefore UV filters (e.g., HST/F225W, HST/F275W, and HST/F336W) are the closest to the peak of the SED and the most sensitive to the SED slope. Additionally, the UV coverage provides the necessary constraints to break the \teff\ - dust degeneracies when combined with IR coverage.  To quantify the effects on SED fitting, we start with the seven-filter fit and remove a UV filter from bluest to reddest for each run, until F475W (optical) becomes the bluest filter. We compare each resulting \beast\ output to the simulated input parameters in Figure \ref{test}, and discuss the results here:

\begin{figure*}[ht!]
\centering
\includegraphics[width=\textwidth]{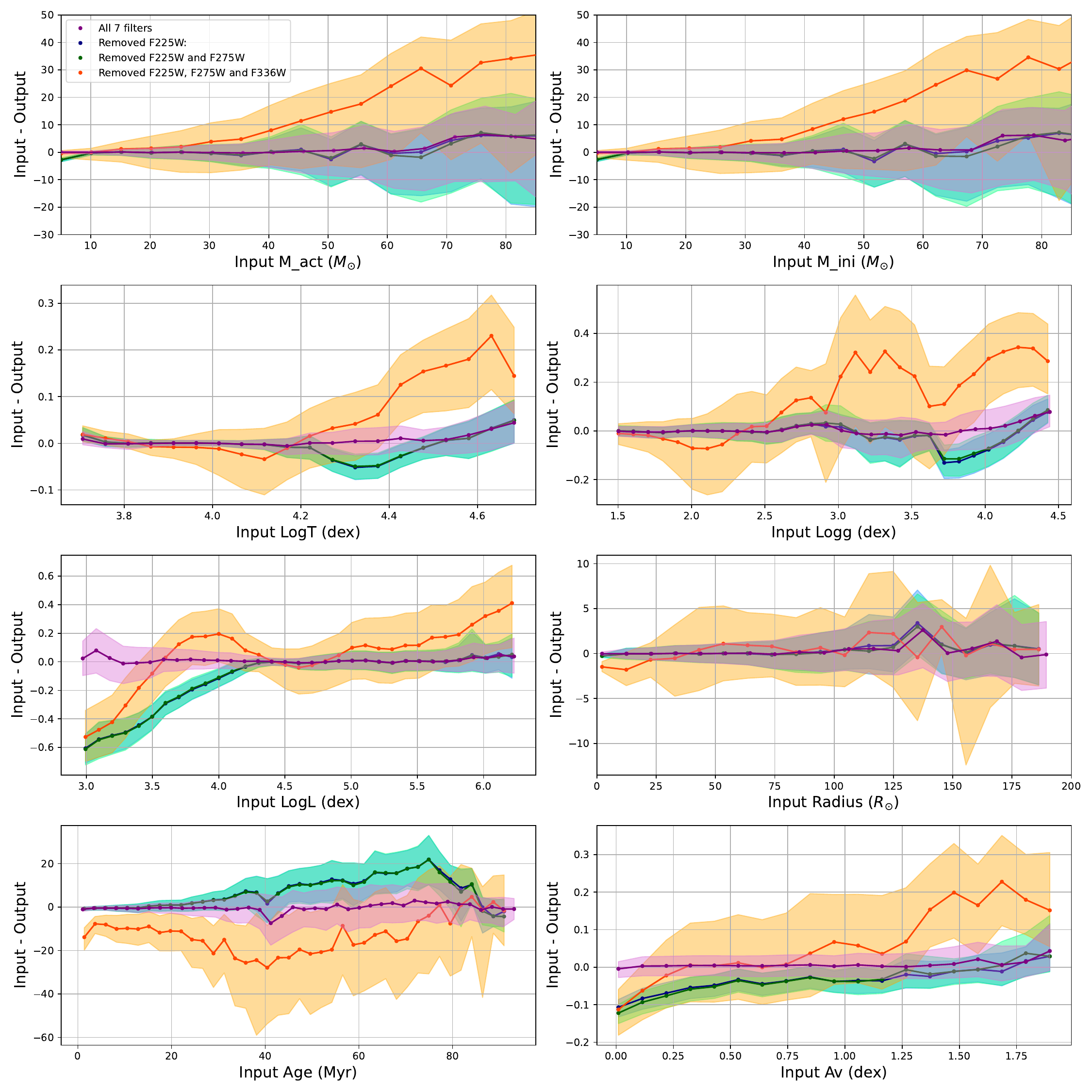} 
 \caption{ The median $\pm 1 \sigma$ difference between the \beast\ output (Output) and simulated parameters (Input) for 45,000 stars. We compare the difference as a function of removing a filter in the NUV until we have only optical and IR bands. For single stars, we recover the parameters ($M_{\rm act}$,$M_{\rm ini}$,\logg,$\log(L)$,\av) for massive stars up to 55 \msun\ consistently. Additionally, we find that at the hot end of $\log(T)$, the \beast\ starts to underestimate the temperature, which is unsurprising, as the hotter temperature pushes the peak of the SED into a regime that the NUV-Optical photometry is less sensitive to. The \beast\ fitting struggles to recover the true parameters as blue filters get removed. In particular, the \beast\ underestimates mass, as noted by \citet{Lindberg25}. When no UV filters are available (No F225W, F275W, F336W), recovering the properties of massive stars is not realistic.} \label{test}
\end{figure*}

\begin{itemize}
    \item \textbf{Masses: $M_{\rm act}$ and $M_{\rm ini}$}: 
    For a single star with seven filter coverage, the \beast\ can recover the input masses within $1 \sigma$ for stars up to 55~\msun, with uncertainties slightly increasing with increasing mass. For stars M $<55$~\msun we find a mean absolute error (MAE) of $\sim 0.7  \pm  1.6 $ \msun, whereas for M $>55$~\msun the MAE increases to $\sim 9 \pm 11 $. The initial mass and actual mass reported MAEs are of similar order for each comparison; hence, we report only one value. For the stars with M $<55$~\msun, we do find only a slight decrease in accuracy between the \beast\ fits that do not include F225W (MAE $\sim1.9 \pm 4.5$~\msun), F225W and F275W (MAE $\sim1.9 \pm 4.5$~\msun) compared to having all three UV filters. While removing F225W and F275W generally leads to larger uncertainties, overall the mass is similarly well recovered by only one UV filter for M $<55$~\msun. 
    However, as we remove all the UV data, we observe a worsening in the recovery of the input; in fact, the \beast underestimates the mass. This is likely a result of sampling the initial stellar mass in the \beast\ fitting from a Kroupa IMF \citep{Kroupa01}. For M $<30$~\msun, we find a MAE $\sim 2.6 \pm 3.2$~\msun, however, for M $>30$~\msun, we find a MAE of $\sim 12.9 \pm 14.2$~\msun. This suggests that a single UV filter is sufficient to recover masses from SED fitting within $\sim2$~\msun. Including HST/F225W in the SED fit recovers masses to within $<1$~\msun accuracy. We caution the reader that the reported masses are model-dependent. While this effect may be less severe at masses below M $<30$~\msun \citep{Gull22}, for higher masses, larger uncertainties persist in stellar evolution that may impact the stellar evolutionary models from which the \beast\ infers the mass.  
    
    \item \textbf{Temperature: $\log(T)$}: The temperature is accurately recovered (MAE: $\sim 650 \pm 970$~K). The sole exception is the hot temperature end, where we observe a slight underestimation of the hottest temperature ($\log(T) > 4.5$, MAE: $\sim 2000 \pm 1800$~K). This divergence is unsurprising, as even F225W will become less sensitive to the SED peak, which will shift to bluer wavelengths. Removing F275W and F336W leads to an underestimation of the temperature between $\log(T) \sim 4.2$ and  $\log(T) \sim 4.5$~dex, while at temperatures above $\log(T) \sim 4.5$~dex we find a similar trend previously described. For $4.2$~dex $< \log(T) < 4.5$~dex, we find a MAE of $\sim 3300 \pm 5500$~K). The underestimation of temperature between $\sim 4.2$~dex and $\sim 4.5$~dex is predominantly driven by stars between $5$~\msun and $10$~\msun, when these stars are removed, the MAE is $1100 \pm 1900$~dex. Most likely this is due to the switch between the underlying stellar atmosphere models in this regime by the \beast\ from the Castelli $\&$ Kurucz \citep{ck04} models to the TLUSTY \citep{lanz03,Lanz07} models. 
    As we remove all the UV filters, two trends happen. First, the \beast\ overestimates $\log(T) \sim 3.8$ and  $\log(T) \sim 4.2$; additionally, it underestimates temperatures higher than $\log(T) \sim 4.2$. Two effects are likely at play here, for one, the \beast\ can no longer break the \teff-dust degeneracy. Secondly, the optical bands are relatively insensitive to the hottest temperature because they are farther from the SED peak. 

    \item \textbf{Surface Gravity: \logg}: For surface gravity, the \beast\ output compared to the input yields a MAE of $0.03 \pm 0.04$~dex. A slight deviation from accurately recovering surface gravity is observed at \logg$>4.5$, where the MAE increases to $0.07 \pm 0.04$~dex. Overall, this suggests an acceptable recovery of surface gravity. Again, we find only a minimal difference between removing F225W and removing both F225W and F275W. Below \logg$ \sim 3.5$~dex, we find that surface gravity is recovered as well as with all bands included (MAE: $0.03 \pm 0.04$~dex). An overestimation of \logg is observed between \logg$\sim 3.5$~dex and $\sim 4.2$~dex, yielding a MAE of $0.13 \pm 0.14$~dex. Once again, if we remove stars between $5$~\msun and $10$~\msun in the temperature range from $\log(T) \sim 4.2-4.5$~dex from the sample, we improve the accuracy (MAE:$0.07 \pm 0.07$). The loss of accuracy here is likely due to the \beast\ switching between the two atmosphere models, which will yield slightly different stellar parameters. 
    The accuracy of the optical-only output recovery is relatively unaffected by removing these stars, hinting that the large scatter and underestimation of the \logg values (MAE: $0.18 \pm 0.26$) is not due to this transition between models. Instead, the underestimation of the \logg values is likely driven by the mass prior. Without the UV constraints, the \beast\ attempts to fit lower masses while increasing the surface gravity to recover the SED shape.   
    
    \item \textbf{Luminosity: $\log(L)$}:
    The input luminosity is well recovered by \beast\ when 7-band coverage is available (MAE: $0.04 \pm 0.06$~dex). 
    When the F225W and F275W bands are removed, the \beast\ overestimates the luminosity of stars with $\log(L)< 4.1$. Investigating the stars that fall in this regime, we find that the \beast\ is overestimating \av values in this regime (mean $-0.26$~dex). If we remove the stars known to cause issues for temperature and surface gravity due to their closeness to the stellar atmosphere grid transition, we find that this trend disappears. 
    When removing all UV bands, the \beast\ consistently underestimates the input luminosity for stars with  $3.6 < \log(L) < 4.3 $ and  $4.7 < \log(L)$. The first bump is driven by lower mass massive stars ($5-15$~\msun). The high-luminosity bump is dominated by young ($<17$~Myr) stars with masses $>30$~\msun, which appear to be driven to lower luminosities by the mass prior. The scatter increases from a standard deviation of $\sim0.05$~dex to a standard deviation of $\sim0.24$~dex. Without the UV, constraining the luminosity of the massive stars is not reliable.
    
    \item \textbf{Radius: $Radius$}:
    For radius, we see an increase of scatter as we remove bands; the standard deviation increases from $\sim 0.7~ R_{\odot}$ when at least one UV band is present to $\sim 2.5~ R_{\odot}$ when none are used. No clear trends present themselves as the MAE are $0.4 \pm 0.5~ R_{\odot}$ and $2.3 \pm 3.0~R_{\odot}$, respectively. 
  
    \item \textbf{Age: $Age$}:
     We find that the \beast\ recovers the ages of most stars (MAE: $2.5 \pm 8.9$~Myr), with the exception of stars around $\sim 40-45$~Myr, even when using all 7 filters.
     As we remove the UV bands, the \beast\ overestimates the age of the stars. Here again, we find that the overestimation is driven by stars between $5-10$~\msun. As we remove these stars, the MAE drops from $5.5 \pm 14.0$~Myr to $1.2 \pm 2.0$~Myr. Removing these stars from the fit with no UV coverage, however, does not eliminate the overall underestimation of the ages. Again, this is ultimately driven by the mass prior, which will make the \beast\ opt for a more evolved, yet less massive, star to match the brightness of the simulated SED. The resulting MAE is $18.5 \pm 60$~Myr, indicating a significant deviation from the true value and a large scatter. Without the UV, constraining the luminosity of the massive stars is not reliable.

    \item \textbf{Extinction: \av}:
    For extinction, we find that input and output are overall consistent (MAE:$0.02 \pm 0.02$~dex). Removing F225W or F225W and F275W does lead to an overestimation of the dust. The tail between \av $\sim0.0 - 0.5$ is predominantly driven by the $5-10$ \msun star, causing issues for all parameters. However, even when removing those, the mean error lies around $\sim -0.02$~dex
    Without UV filters, the uncertainty on the \av\ increases significantly (mean standard deviation:  $\sim -0.10$~dex), and is overestimated for low \av\ and underestimated for high \av. 
    
\end{itemize}

Overall, the results suggest that even one UV filter can significantly improve the recovery of stellar parameters for single massive stars up to $\sim 55$ \msun.

\subsection{The effects of fixing metallicity, distance, and the extinction curve}\label{sec:fix}
In the analysis of our papers, we fix metallicity($Z$), distance, and the extinction curve (Rv) when fitting to data. To quantify the effects on SED fitting, we generate a simulated data set with distance in the range of $25.56-25.81 \rm ~mag$, $Z$ in the range of $3\%-11\%$~\zsun, and $R_V$ in the range of $2.06 - 3.20$. These ranges are intended to be more generous than the physical scenario. We then perform the \beast~fitting assuming fixed $Z$, distance, and $R_V$. 
We compare each resulting \beast\ output with the simulated input parameters and divide the difference by the uncertainty to quantify the significance of the deviations (See Fig. \ref{fixtest}). The results of this significance test are discussed here:

\begin{figure*}[ht!]
\centering
\includegraphics[width=\textwidth]{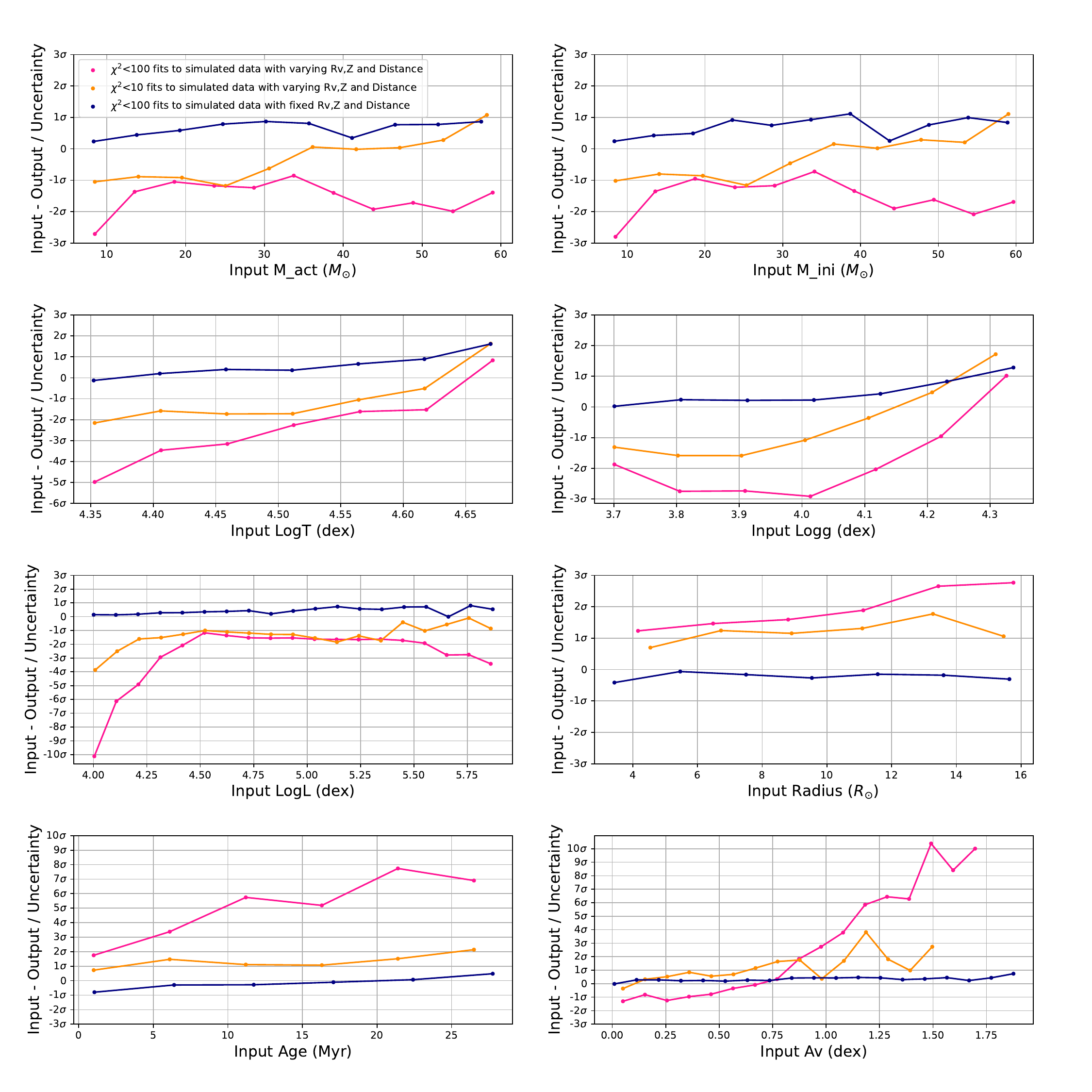} 
 \caption{Median difference between the \beast\ output and simulated input parameters divided by 84th percentile minus 16th percentile divided by 2 (Uncertainty) for 8546 stars with $\chi^2<10$ (orange) and 23939 stars with $\chi^2<100$ (pink). We compare the effects of fixing $R_V$, $Z$, and distance when fitting SEDs to simulated data in which these parameters are not fixed. We also plot the results of fitting to simulated data with fixed $R_V$, $Z$, and distance (blue), where nearly all data points are consistent within $\sim1~\sigma$. For the $\chi^2<10$ fits, the results lie generally within $\sim 2~\sigma$, suggesting that deviations from the expected values are mostly due to noise.  For the $\chi^2<100$ fits, most parameters are consistent within $\sim~3~\sigma$. However, the low value end of $\log(L)$, the high value end of $A_V$ and the ages are clearly $>3~\sigma$. The low value end of $\log(L)$, which also drive the divergence in the ages, is most likely caused by the switch between atmosphere models in this regime. Removing $\log(L)<4.3$ from the sample, shifts the majority of the ages in to the statistically consistent regime. For our analysis, these results suggest that fixing $R_V$, $Z$, and distance,  does not introduce significant trends in the parameters and parameter range of interest.} \label{fixtest}
\end{figure*}

%Exceptions are the low value end of $\log(L)$ and the high value end of $A_V$, indicating that they are close to statistically consistent but with reduced sensitivity. In physical units, for the $\chi^2<100$ sample this deviation is $<2.5~ \rm Myr$ for ages up to $\sim 15~ \rm Myr$ and  $<5~ \rm Myr$ up to $\sim 20~ \rm Myr$, despite the rather large $\rm Z- \rm score$.

\begin{itemize}
    \item \textbf{Masses: $M_{\rm act}$ and $M_{\rm ini}$}: $M_{\rm act}$ and $M_{\rm ini}$ yield $|\rm Z-\rm scores|$ ((Input$-$Output) $/$ Uncertainty) within $3~\sigma$, with the majority of values below $2~\sigma$. For the $\chi^2<10$ sample, we find that the $|\rm Z-\rm scores|$  lie around or below $1~\sigma$. For the $\chi^2<100$ sample, we find that majority of the $|\rm Z-\rm scores|$  lie between $1~\sigma$ and $2~\sigma$. In both samples, fixing $R_V$, $Z$, and the distance yields slightly higher masses at the low-mass end. In physical units, we find that for $\chi^2<10$ and masses $<40$~\msun the deviation ranges from $\sim1$~\msun at the low-mass end to $\sim5$~\msun at the high-mass end. In general, these results suggest that $M_{\rm act}$ and $M_{\rm ini}$ are relatively unaffected by the choice to fix $R_V$, $Z$, and distance relative to the uncertainties of the fitting itself, when choosing $\chi^2$ to restrict the data analyzed.
        
    \item \textbf{Temperature: $\log(T)$}: For temperature, we find that the $\chi^2<10$ sample yield $|\rm Z-\rm scores|$  within $2~\sigma$ and the $\chi^2<100$ sample yield $|\rm Z-\rm scores|$  within $3~\sigma$ for temperatures $\log(T)>4.45$. They appear to exhibit a temperature-dependent trend: low temperatures yield higher $\sigma$ values, whereas high temperatures yield lower $\sigma$ values. For the $\chi^2<100$ sample, temperatures below $\log(T)\sim4.45$ show $|\rm Z-\rm scores|$  larger than $3~\sigma$ suggesting a significant deviation from the expected outcome. We note that this temperature range coincides with the range in which the switch between stellar atmosphere models occurs in the \beast (See Appendix \ref{sec:UV}). In this case despite having 7-filter coverage the treatment of $R_V$, $Z$, and distance may result in similar uncertainties. In physical units the temperature deviation ranges from $1618$~K to $3687$~K. Although our analysis uses temperature values of SED fits with $\chi^2<100$, we note that the majority of the stars discusses have temperatures $\log(T)>4.4$; only 40 stars have $\log(T)<4.4$ in the $\chi^2<100$ regime. 
    These results suggest that $\log(T)$ is relatively unaffected by our fitting choices for the stars of interest in this analysis.

    \item \textbf{Surface Gravity: \logg}: For surface gravity, we find that the $\chi^2<10$ sample yields $|\rm Z-\rm scores|$  within $2~\sigma$ and the $\chi^2<100$ sample yield $|\rm Z-\rm scores|$  within $3~\sigma$. For lower surface gravity, the \beast\ overestimates surface gravity, and at higher surface gravity, the \beast\ underestimates surface gravity. In physical units, the deviations remain within $\pm0.1 \rm~dex$, and are relatively unaffected by our fitting choices in this regime.
    
    \item \textbf{Luminosity: $\log(L)$}: The input luminosity yields $|\rm Z-\rm scores|$  within $2~\sigma$ for $\log(L) > 4.2$~dex, clustering around $1~\sigma$ for the $\chi^2<10$. The $\chi^2<100$ sample deviates more significantly for $\log(L) >4.3$~dex. This deviation appears to be predominantly driven by the switch between underlying models as described in the previous section, combined with the treatment of $R_V$, $Z$, and distance. The \beast overestimates the luminosity compared to the input sample on the lower luminosity end. In physical units, this deviation ranges from $\sim 0.3$ dex to $\sim 0.1$ dex. Furthermore, the simulated stars with $\log(L) < 4.25$~dex have masses with $M_{star}<9$\msun. We note that this test likely overestimates the divergence, as we explored a generous range of values for $R_V$, $Z$, and distance. In terms of our analysis, this only impacts Fig. \ref{HR}, since we otherwise do not rely on $\log(L)$. While the deviation is likely smaller in the actual fitting, caution should be taken when the luminosity results with $\log(L) < 4.25$~dex from Table \ref{tab:beast} are used in the future. 
    
    \item \textbf{Radius: $Radius$}: For radius, we find that both the $\chi^2<10$ and the $\chi^2<100$ sample yield $|\rm Z-\rm scores|$  within $2~\sigma$ and $3~\sigma$, respectively. The radius is typically underestimated by \beast; no other trend is apparent. In physical units, the deviations are $<1$~\rsun and suggest that our fits are relatively unaffected by the fitting choices.
  
    \item \textbf{Age: $Age$}: The input ages yield $|\rm Z-\rm scores|$  within $\sim2~\sigma$ for the entire $\chi^2 <10$ sample. For the  $\chi^2 <100$ sample, only ages $<5~ \rm Myr$ appear to be below $\sim3~\sigma$. For both samples, the \beast underestimates the ages. This is perhaps not surprising since distance and metallicity will directly impact the choice of isochrone and therefore impact ages the most. Upon further investigation it is also evident that the deviation is primarily driven by stars with $\log(L) < 4.25$~dex. Again, this is likely a reflection of the model transition zone manifested in the fitting. If we remove stars with $\log(L) < 4.25$~dex, the majority of the $|\rm Z-\rm scores|$  fall below $\sim3~\sigma$ and all fall below $\sim4~\sigma$. For our analysis this in particular could impact the estimation of velocities for the runaway candidates. However, all six candidates have luminosities $\log(L) > 4.3$~dex, and should therefore not be impacted. In physical units, for the $\chi^2<100$ sample this deviation is $<2.5~ \rm Myr$ for ages up to $\sim 15~ \rm Myr$,  $<5~ \rm Myr$ up to $\sim 20~ \rm Myr$ and $<10~ \rm Myr$ up to $\sim 30~ \rm Myr$. While the $\chi^2 <10$ sample is statistically robust for our analysis, interpreting results in the $\chi^2<100$ sample requires attention to the specific parameter regime in which the stars reside.
        
    \item \textbf{Extinction: \av}: \av is also significantly impacted by our choice of fixing $R_V$, $Z$, and distance. In particular, fixing $R_V$ when the underlying $R_V$ is varying causes the rather significant deviation at large \av. We note that for the $\chi^2<10$ sample, the $|\rm Z-\rm scores|$ remain within $2~\sigma$ up to \av$=1.00$, and then remains within $\sim 3~\sigma$. For the $\chi^2<100$ sample, the similarly stay within $2~\sigma$ up to \av$=1.00$ and then diverges increasingly with increasing \av. However, we vary $R_V$ substantially in our simulated data, which is unlikely to represent the intrinsic variation observed in Sextans A. In our analysis, this affects only Fig. \ref{uvmap}, where we use the attenuated SED to generate the simulated GALEX and UVEX maps. 
    
\end{itemize}

Overall, the results suggest that fixing $R_V$, $Z$, and the distance does not significantly affect our analysis.

\section{Limitations of the \beast\ in regards to massive binaries}\label{sec:singletracks}

\begin{figure}[ht!]
\includegraphics[width=\columnwidth]{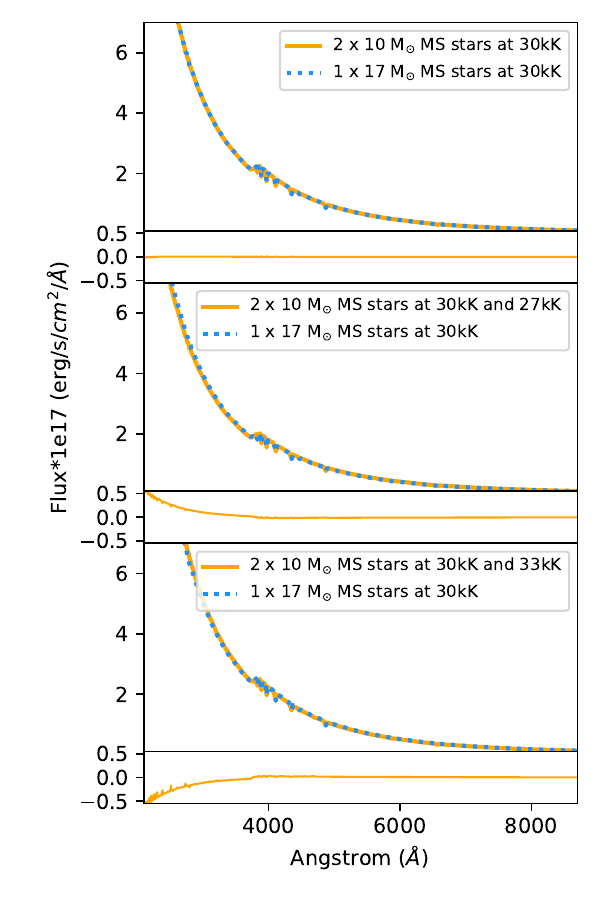}
 \caption{We show synthetic spectra of massive stars and binaries. The top panel shows the spectra of a twin binary system consisting of two 10 \msun\ MS stars, each with T$_{eff} = 30kK$ (orange). We overplot the spectra of a single 17 \msun\ MS star with T$_{eff} = 30kK$ (blue). The two spectra are nearly indistinguishable; inspecting the residuals, we see little difference between the two SEDs. Only if we zoomed in on the spectral features would we see minimal differences in line depth, which is unlikely to be captured by photometry alone. The middle and lower panels repeat the exercise, but the companion is once cooler (middle) and once hotter (bottom). We note that in the optical ($\lambda>4000$ Angstrom), there is nearly no difference between the binary and the single star. However, the UV shows a deviation not covered by our HST photometry and will yield a suboptimal $\chi^2$.} \label{spectracomp}
\end{figure}

Binaries and binary interaction are omnipresent among massive stars. Observations and predictions suggest that the binary fraction of stars with $M_{*}> 8$~\msun\ ranges from $85 \%$ to $100 \%$ \citep{Sana2012, Moe17, Offner23}. 
Yet for our panchromatic photometry analysis of Sextans A, the stellar evolution and atmosphere models used throughout this study are single-star models. Given what we know about massive stars and binarity, it is clear that these will not be adequate models or assumptions for all stars. We briefly discuss here which shortcomings, in particular, will be problematic, as they will present themselves as stars with reliable $\chi^2$ fits. 

For the pre-interaction binaries, we can encounter four fitting scenarios. First, for some systems, the primary will still be photometrically so dominant that the \beast\ fit with the uncertainties on our measurements will adequately capture the primary. 

Second, twin binaries, expected to make up $3\%-14\%$ \citep{Moe17}, will likely yield a reliable $\chi^2$ value for the \beast\ fit, since the SED will be indistinguishable from a star with similar temperature but larger mass (See Figure \ref{spectracomp}). In that case, the \beast\ will mistake the twin system as a more massive single star. 

Third, in the case of eclipsing binaries, the \beast\ will yield a poor $\chi^2$ value ($>100$ in most cases) for short-period binaries, since the individual data points are likely not measured at the same step of phase space the variations for detectable eclipsing binaries ($\sim 1-5\%$) will lead to large enough divergence from a single star SED that the \beast\ will not recover it. 

The last scenario concerns pre-interaction binary configurations that have not yet been covered. The more the stars in the system diverge in type, the higher the $\chi^2$ value will be (See Fig. \ref{spectracomp}). Therefore, a subset will likely still fall in the plausible $\chi^2$ regime ($<10$), while a subgroup will fall in the plausible SED fits regime ($<100$). The remaining will fall within the bad $\chi^2$ regime. 

Future studies may want to consider fitting two stellar SEDs to systems falling within the $10<$ $\chi^2$ $<100$ regime to determine if a better fit can be recovered. For all of the aforementioned binary scenarios, the masses of the stars are overestimated in the worst-case scenario. 

For the post-interaction binaries and products, a subset will yield poor $\chi^2$ values (e.g., classical Be stars with a disk in line of sight, stripped stars). However, a subgroup of post-interaction stars will appear as single stars, yielding a plausible SED fit. In that case, the stellar parameters may not accurately reflect the star's true parameters. 
Post-interaction mass gainers can appear over-luminous for their mass. In contrast, mass donors may appear under-luminous for their mass. In both cases, fitting a single-star atmosphere model will yield erroneous stellar parameters, even when the SED fit is robust. To first degree, the temperature is likely the only reliable parameter in this case. These SEDs are perhaps the only ones where fitting current single-star atmosphere models misses the majority of the underlying physics.

\acknowledgments
Maude Gull is a Carnegie-Caltech Brinson Fellow supported by The Brinson Foundation. MG also acknowledges support of the UC Berkeley Cranor Fellowship, and the Schweizerische Studienstiftung for this work. MG thanks Peter Senchyna for helpful discussion. MG also thanks the UC Berkeley Shining Lights program for motivational support.

Support for this work was provided by NASA through grants GO-15275, GO-15921, GO-16149, GO-16162, GO-16717, GO-17438, AR-15056, AR-16120, AR-17026, HST-HF2-51457.001-A, and JWST-DD-1334 from the Space Telescope Science Institute, which is operated by AURA, Inc., under NASA contract NAS5-26555.

This research used the Savio computational cluster resource provided by the Berkeley Research Computing program at the University of California, Berkeley (supported by the UC Berkeley Chancellor, Vice Chancellor for Research, and Chief Information Officer).

Grammarly (non-AI version) and Microsoft Copilot were used to check grammar, spelling, and sentence structure of some individual paragraphs and sentences. They were not used to generate text. 

This work is based on photometric observations made with the NASA/ESA Hubble Space Telescope, obtained from the data archive at the Space Telescope Science Institute. STScI is operated by the Association of Universities for Research in Astronomy, Inc. under NASA contract NAS 5-26555. The HST data presented in this article can be obtained from the Mikulski Archive for Space Telescopes (MAST) at the Space Telescope Science Institute. 

%10.26131/IRSA414 LVL

A survey paper will provide relevant details for the future LUVIT data release and a full dataset DOI (Boyer et al., in prep).

This work made extensive use of NASA's Astrophysics Data System Bibliographic Services.
\facility{NED}
\software{\texttt{astropy} \citep{astropy:2013, astropy:2018, astropy:2022}, \beast\ \citep{Gordon16},\texttt{DOLPHOT} \citep{DOLPHOT},
\texttt{matplotlib} \citep{matplotlib}, \texttt{numpy} \citep{numpy}, \texttt{pyphot}}

\bibliography{mybib}

\end{document}